\documentstyle[12pt,epsfig]{article}
\input epsf
\newcommand{\be}{\begin{equation}}
\newcommand{\ee}{\end{equation}}

\def\fun#1#2{\lower3.6pt\vbox{\baselineskip0pt\lineskip.9pt
\ialign{$\mathsurround=0pt#1\hfil##\hfil$\crcr#2\crcr\sim\crcr}}}

\begin{document}

\title{Scalar mesons and low-mass sigma: Does the $\sigma$ reveal
the confinement singularity? }

\author{V.V. Anisovich  }

\date{\today}

\maketitle

\begin{abstract}
I present a short review of the current understanding of the scalar
meson sector, with special attention to the problem of the low-mass
$\sigma$.
The dispersion relation
$N/D$-method used for the restoration of the low-energy $\pi\pi$
$(IJ^{PC}=00^{++})$-wave amplitude is discussed. The low-energy
$\pi\pi$ amplitude was determined
 from the data
in the energy region 280--500 MeV and it was sewn  with
the previously found $K$-matrix solution for the region 450--1950 MeV.
The $N/D$-amplitude has a pole on the second sheet of the complex-$s$
plane, near the $\pi\pi$ threshold at $\sqrt s \simeq\ 430-i325$ MeV,
that corresponds to the low-mass $\sigma$-meson. I discuss the
hypothesis that this pole may be related
to the confinement forces, thus being the eyewitness of confinement.
\end{abstract}

\section*{Introduction}
In the region of strong interactions, the fundamental objects of QCD,
quarks and gluons, may substantially change their characteristics as
compared to those at small distances. Phenomenological investigations,
as well as theoretical estimates, show that in the strong interaction
region we deal with constituent quarks and effective
massive gluons (for more detail see, for example, \cite{book}
and references therein). In the paper \cite{Gribov},
V.N. Gribov drew the attention of the reader to the fact that
strong interactions, especially interactions caused by the confinement
forces, may form particles which are different from the standard
(non-exotic) mesons and baryons. First of all, it concerned  the  sector
of scalar mesons. In the present paper, I discuss the hypothesis of
whether the low-mass $\sigma$ meson, if it exists, may point to the
existence of the amplitude singularities related to confinement.

\section{Scalar mesons}

In this Section, I briefly present the results of  the
K-matrix analysis for the $(J^P=0^+)$-wave resonances and the nonet
classification of these states. Special attention is paid to the status
of $f_0(980)$, $a_0(980)$ and $ f_0(1300)$.

\subsection{ \boldmath$K$-matrix analysis at $\sqrt s > 450$ MeV}

The experimental data on meson spectra accumulated by several
groups, see
\cite{GAMS,GAMS-eta,BNL,C-M,cbc,cbc_new,BNL-new},
provided us with a good basis for the study of the
$(IJ^{PC}=00^{++})$-wave;
the $K$-matrix analysis of the reactions $\pi\pi \to
\pi\pi$, $K\bar K$, $\eta\eta$, $\eta\eta'$, $\pi\pi\pi\pi$ was carried
out over the mass range 450--1900 MeV in a set of papers
\cite{APS,YF,K}.   Also
the $K$-matrix analysis has been performed for
the waves $10^{+}$ \cite{YF} and $\frac12 0^{+}$ \cite{AlexSar} ,
thus making it possible to establish the $q\bar q$ systematic of
scalars for $1^3P_0 q\bar q$ and $2^3P_0 q\bar q$ multiplets.

  In the scalar--isoscalar sector, $IJ^{PC}=00^{++}$, the following
  states are seen:
\begin{eqnarray}
00^{++}: \quad && f_0(980),\ f_0(1300),\ f_0(1500)\ , \nonumber\\
&& f_0 (1200-1600),\ f_0(1750)\ .
\label{1'}
\end{eqnarray}
The location of corresponding poles on the complex-mass plane
($\sqrt s \equiv M$) is shown in Fig. 1.

\begin{figure}
%Fig. 1
\centerline{\epsfig{file=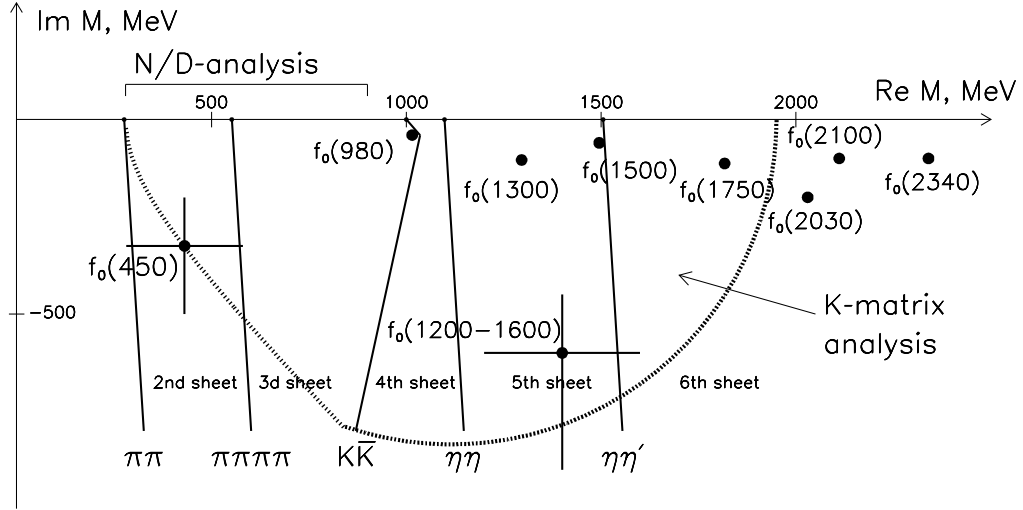,width=14cm}}
\caption{ Complex $M$ plane ($M\equiv \sqrt s $) in the
$(IJ^{PC}=00^{++})$ sector. Dashed line encircles the part of the plane
where the $K$-matrix analysis reconstructs the analytical $K$-matrix
amplitude: in this area the poles corresponding to the resonances
$f_0(980)$, $f_0(1300)$, $f_0(1500)$,
$f_0(1750)$ and the broad state $f_0(1200-1600)$ are
located. On the border of this area
 the light $\sigma$-meson denoted as
$f_0(450)$ is shown (the position of the pole corresponds to that found
 in the $N/D$ method). Beyond the $K$-matrix analysis area, there
 are resonances
$f_0(2030),f_0(2100),f_0(2340)$. }
\end{figure}

The $f_0(980)$  is a well-known  resonance, its properties and its
nature are intensively discussed during several decades.
In the compilation \cite{PDG}, the $f_0(1300)$ is denoted  as
$f_0(1370)$, however, following the most accurate
determination, its mass is near 1300 MeV -- so
the notation $f_0(1300)$ is used here. The $f_0(1500)$ resonance
had been discovered in
\cite{f1500,f1500cb,f1500PR},
now it is a well-established state.  A few years ago, there existed a
strong belief that in the region around 1700 MeV  a
comparatively narrow state $f_J(1710)$
was present, with $J=0$ or 2. But the
$K$-matrix analysis \cite{APS,YF,K}
pointed to the resonance
$f_0(1750)$, with the width $\Gamma\sim 140-300$ MeV: the uncertainty
in the definition of its width is due to a poor knowledge
of the $\pi\pi\pi\pi$ channel in this mass range and, correspondingly,
to the two available solutions,   $\Gamma\sim 140\,$MeV and
$\Gamma\sim 300\,$MeV.

The broad state $f_0(1200-1600)$,
with a half-width $500-900$ MeV, is definitely needed for the $K$-matrix
analysis. In \cite{YF}, this  state had been denoted as
$f_0(1530^{+90}_{-250})$: a large error in the definition of the mass
is due to the remoteness of the pole from the real axis (physical
region) as well as to the existence of several solutions given by the
$K$-matrix analysis.

The analysis of a large number of reactions,
where the broad state $f_0(1200-1600)$ reveals itself, proved the
validity of the factorisation inherent in the resonance amplitude: near
the pole the amplitude is represented as $g_{in}(s-M^2)^{-1}g_{out}$,
where the universal coupling constants $g_{in}$ and $g_{out}$ depend on
the type of the initial and final states only. A strong production of
the $f_0(1200-1600)$ in various processes allowed us  to fix reliably
these coupling constants.

In the $10^{++}$, $\frac{1}{2}0^{+}$ sectors, the
$K$-matrix analysis \cite{YF,AlexSar} pointed to the presence of the
following resonances:
\begin{eqnarray}
\label{2'}
10^{++}&:&\quad a_0(980),\ a_0(1520)\ , \\
\frac{1}{2}\ 0^+&:& K_0(1415),\ K_0(1820)\ .
\end{eqnarray}
Consideration of the K-matrix bare states argues that the
resonances (\ref{1'}) and (\ref{2'}) are related to the first
and second $q\bar q$ nonets:
\begin{eqnarray}
\label{2''}
n=1:\quad && f_0(980),\ f_0(1300),\ a_0(980),\ K_0(1415)\ ,   \\
n=2:\quad && \ f_0(1500),\ f_0(1750),\ a_0(1520),\  K_0(1820)\ ,
\nonumber
\end{eqnarray}
where $n$ is the radial quantum number of the $q\bar q$ systems; the
broad state $f_0 (1200-1600)$ is the descendant of the scalar glueball.

\subsubsection{The \boldmath$K$-matrix bare states}

A significant trait of the $K$-matrix analysis is that it also gives
us, along with the characteristics of real resonances, the positions of
levels before the onset of the decay channels, i.e. it determines the
bare states. In addition, the $K$-matrix analysis allows one to observe
the transformation of bare states into real resonances.

Let us illustrate the determination of bare states in the K-matrix
technique using a simple example of the one-channel amplitude (for
example, $\pi\pi\to\pi\pi$) with one-resonance state. The amplitude is
written as follows:
\be \label{3'}
 A(s)\ =\ \frac{K(s)}{1-i\rho(s)K(s)}\ ,
\ee
where $\rho(s)=\sqrt {1-4m_\pi^2/s}$ is the phase space factor and
 the
$K$-matrix block contains a pole term related  to the bare state:
\be
\label{4'}
K(s)\ =\ \frac{g^2_{bare}}{m^2_{bare}-s}+f\ .
\ee
One may re-write (\ref{3'}) to the form
\be
\label{5'}
 A(s)\ =\ \frac{g^2_{bare}+(m^2_{bare}-s)f}{m^2_{bare}-s-
i\rho(s)[g^2_{bare}+(m^2_{bare}-s)f]}\ ,
\ee
which, near $s=m^2_{bare}$, turns into the standard Breit-Wigner
expression:
\be
\label{6'}
 A(s)\ =\ \frac{g^2_{bare}}{m^2_{bare}-s-
i\rho(m^2_{bare})g^2_{bare}}\ .
\ee

\begin{figure}[h]
%Fig. 2
\centerline{\epsfig{file=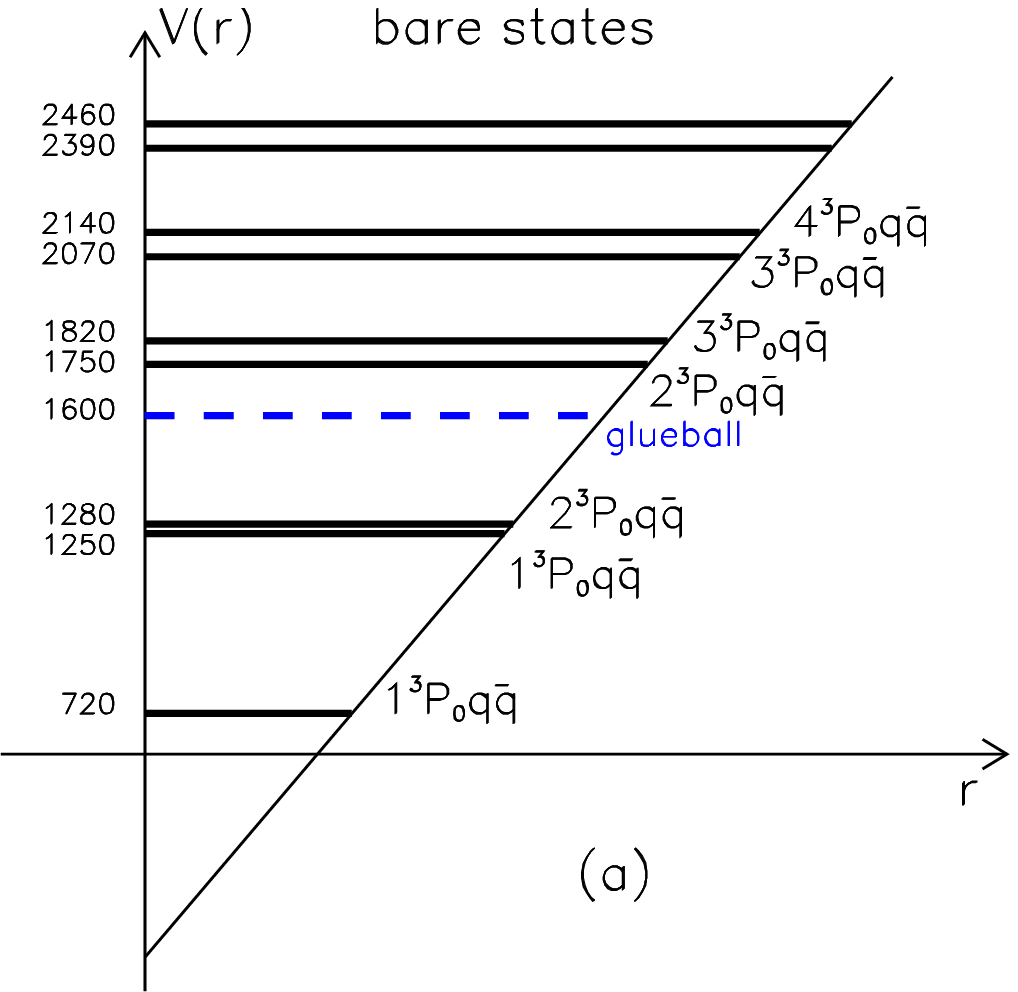,width=8cm}
            \epsfig{file=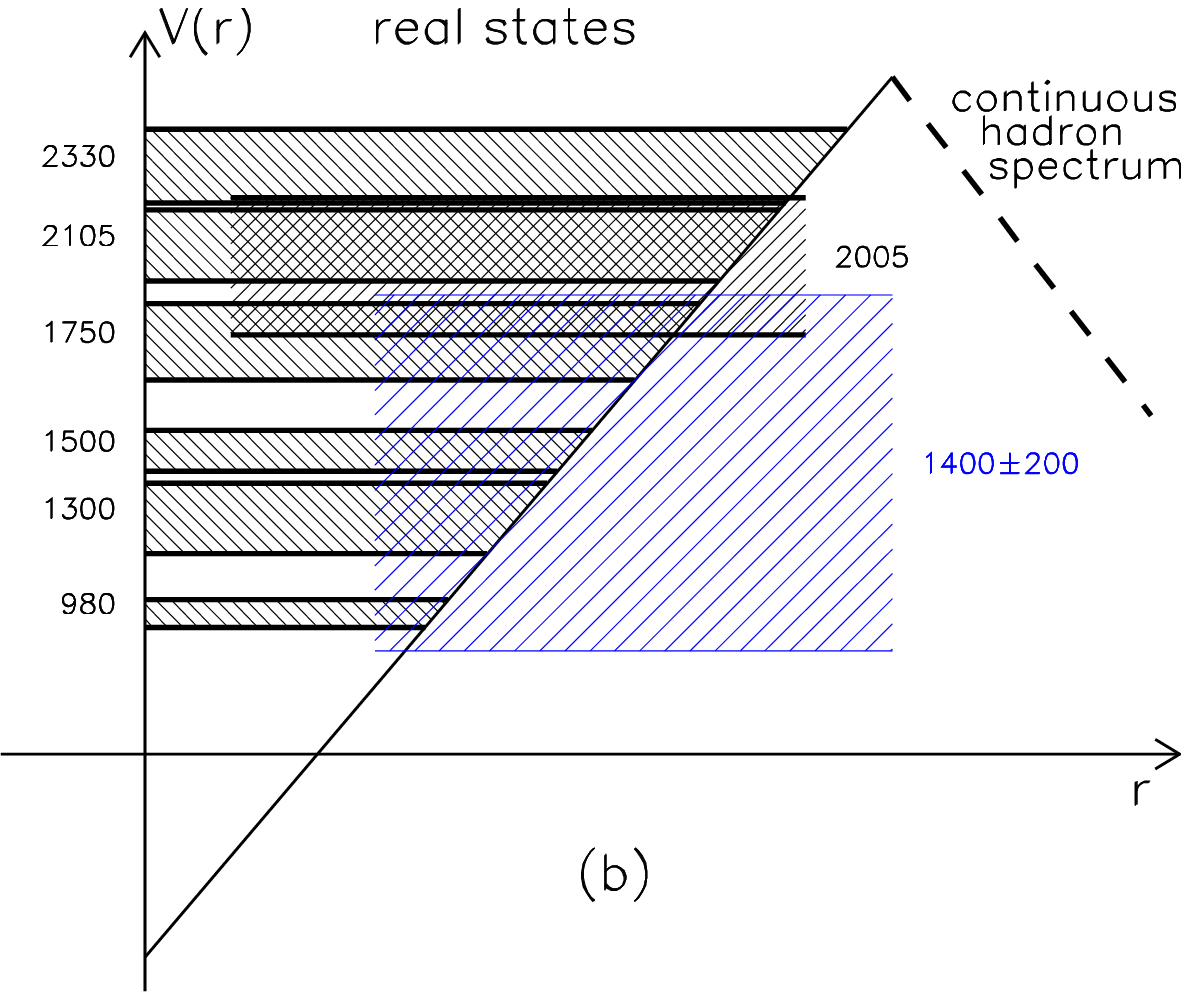,width=8cm}}
\caption{Illustration of the
transform of the bare $00^{++}$-states into real resonances:
 a) bare states are the
levels
in a well with impenetrable wall; b) after the onset of the decay
channels (under-barrier transitions), the stable levels
transform into real resonances. }
\end{figure}

If the imaginary part in the denominator in (\ref{6'})
is neglected, one has
a pole at $s=m^2_{bare}$ corresponding to the stable level in Fig. 2a,
while the pole
in (\ref{5'}) describes the unstable state of Fig. 2b.

Considering scalar resonances
\cite{APS,YF,K,AlexSar},
several coupled channels were analysed,  with the inclusion of several
resonances into the fitting procedure.
This corresponded to the amplitude
represented in the matrix form, $\hat A(s)$, and the matrix metrics
was determined by a number of channels:
\be \label{7'}
\hat A(s) =
\frac{\hat K(s)}{1-i\hat \rho(s)\hat K(s)}\ .
\ee
Here $\hat \rho(s)$
is the diagonal phase space matrix; in \cite{APS,YF,K,AlexSar},
$\hat K(s)$ was
parameterised in the following form:
\be
\label{8'}
 K_{ab}(s)= \sum_n
 \frac{g^{(a)}_{n(bare)}g^{(b)}_{n(bare)}}{m^2_{n(bare)}-s}+f^{(ab)}(s)
 \ .
 \ee
The indices $a,b$ run over different channels, while $n$ refers to the
bare states; $f^{(ab)}(s)$ is a smooth background term.
The positions of amplitude poles are defined by the equality:
\be
\label{9'}
det \left |1-i\hat \rho(s)\hat K(s)\right | =0\ ,
\ee
which determines masses and widths of resonances, while the masses of
bare states are determined by the  K-matrix poles (\ref{8'}).

\begin{figure}[h]
%Fig. 3
\centerline{\epsfig{file=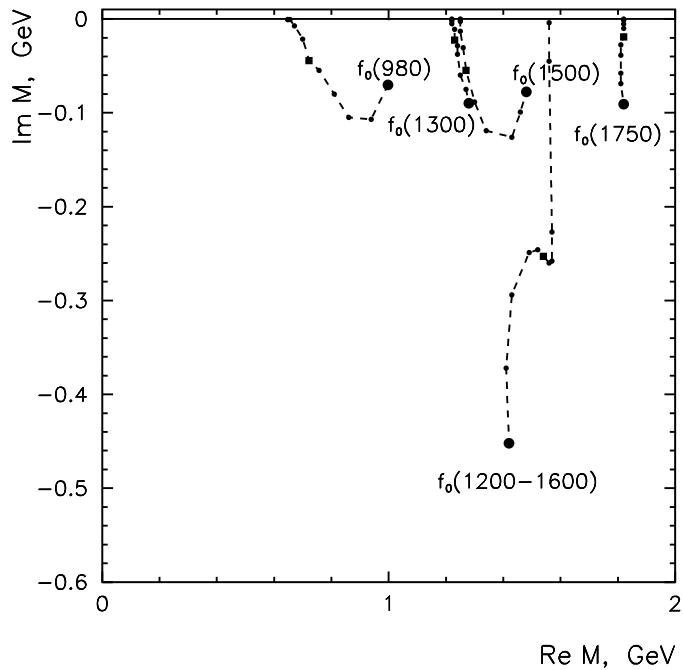,width=10cm}}
\caption{Complex-$M$ plane: trajectories of poles
related to the states $f_0(980)$, $f_0(1300)$, $f_0(1500)$,
$f_0(1750)$, $f_0(1200-1600)$, under a uniform onset of the decay
channels.}
\end{figure}

This evolution of states is illustrated by Fig. 3, where
the trajectories of amplitude poles
in the complex plane are
shown depending on the gradual onset of the decay channels.
Technically, it is not difficult to switch on/off the decay channels
for the  $K$-matrix amplitude:
one should substitute in
the $K$-matrix elements  (\ref{8'}):
\be
g^{(a)}_{n(bare)}
 \to\ \xi_n(x)g^{(a)}_{n(bare)}\ , \qquad f^{ab}\ \to\
\xi_f(x)f^{ab}\ ,
\label{10'}
\ee
where the parameter-functions for  switching on/off the decay
channels, $\xi_n(x)$ and $\xi_f(x)$,  satisfy the
following  constraints:
$\xi_n(0)=\xi_f(0)=0$ and $\xi_n(1)=\xi_f(1)=1$, and $x$ varies
in the interval $0\le x\le 1$. Then, at $x=0$, the amplitude $\hat A$
turns into the $K$-matrix,
 $\hat A(x\to 0)\to\hat K$, and the amplitude poles occur on the
real axis, corresponding to the stable $f^{bare}_0$-states. At $x=1$,
we deal with the real resonance; varying $x$ from $x=0$ to $x=1$ we
observe the movement of poles in the complex  $M$-plane.

\subsubsection{Quark combinatoric
rules for the decay couplings of the quark--antiquark states}

The $K$-matrix amplitude analysis is a good instrument
to perform the $q\bar
q$ nonet classification of mesons in terms of the bare states.
The necessity to use bare states but not real resonances is dictated
by the fact that decay processes produces a
strong mixing of scalar states, for in the transitions $(q\bar q)_1
\to real\; mesons \to (q\bar q)_2$ the
orthogonality of the states
is not preserved.

The decay couplings of the
$q\bar q$-meson or glueball to a pair of mesons are determined
by the planar diagrams with $q\bar q$-pairs
produced by  gluons: these diagrams provide
the leading terms in the 1/N expansion \cite{t'Hooft},
while nonplanar diagrams give the next-to-leading order
contribution.
The examples of
planar diagrams for the decay of quarkonium and gluonium states are
presented in Fig. 4.

\begin{figure}[h]
%Fig. 4
\centerline{\epsfig{file=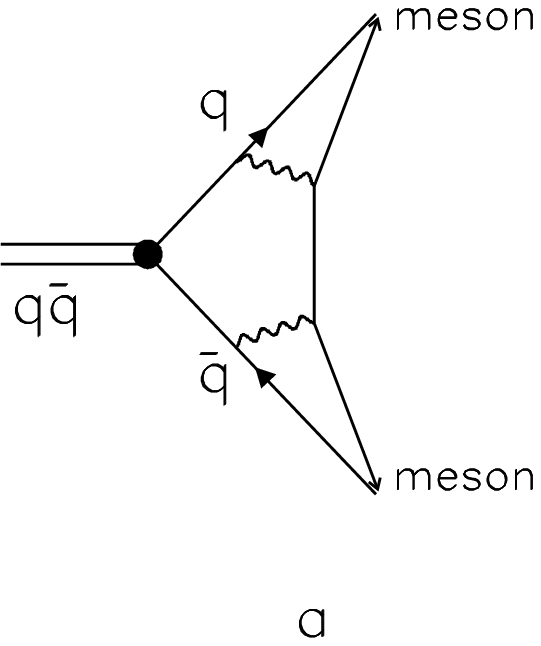,height=5cm}\hspace{0.5cm}
            \epsfig{file=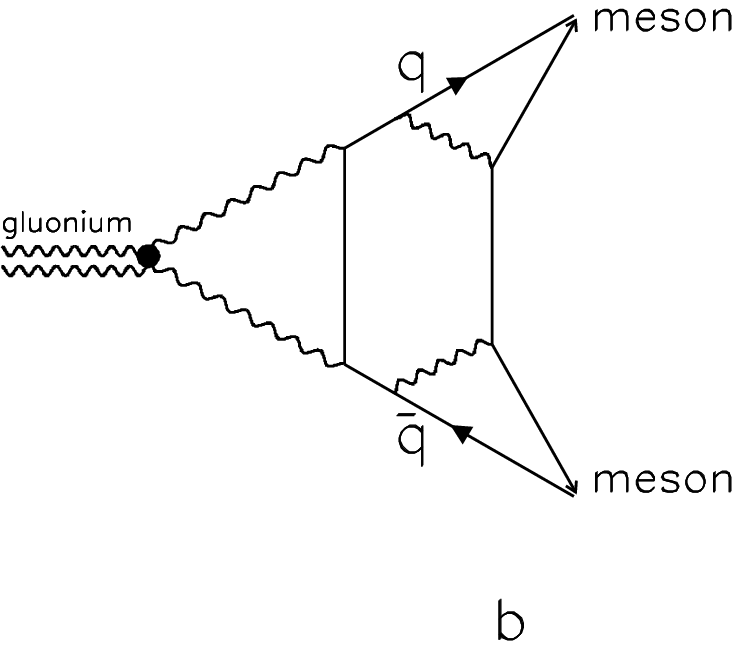,height=5cm}}
%\vspace{0.5cm}
\caption{(a,b) Examples of planar diagrams
responsible for the decay of the  $q\bar q$-state and gluonium into
two $q\bar q$-mesons (leading terms in the $1/N$ expansion).}
\end{figure}

The production of soft $q\bar q$ pairs by
gluons violates flavour symmetry, with the following ratios of the
production probabilities:
\be
u\bar u:d\bar d:s\bar s=1:1:\lambda \; .
\label{lambda}
\ee
The suppression parameter $\lambda$ for the production of strange
quarks varies in the interval $0\leq \lambda\leq 1$. The estimate
made for
high energy collisions gives $\lambda=0.4-0.6$ \cite{book}, while in
hadron decays it was evaluated as $\lambda\simeq 0.8$ \cite{peters}.

In the K-matrix analysis, the quark combinatorial relations are
imposed for the decay couplings of bare states, $g^{(a)}_{n(bare)}$.
The quark combinatoric rules were initially suggested for the high
energy hadron production \cite{as-bf} and then extended for hadronic
$J/\Psi$ decays \cite{wol}. The quark combinatoric relations were used
for the decay couplings of the scalar--isoscalar states
in the analysis of the quark--gluonium content of resonances in
\cite{glp} and later on in \cite{APS,YF,K,AlexSar}.

For the decay couplings of $a_0$  states, the planar diagrams
of the Fig. 4a-type give the following relations:
\be
\label{a0decay}
a_0^-\to K^-K^0:\; \frac{g\sqrt \lambda}{2}\ , \quad
a_0^-\to \eta\pi^-:\; \frac{g \cos\Theta}{\sqrt 2}\ , \quad
 a_0^-\to \eta'\pi^-:\;  \frac{g \sin\Theta}{\sqrt 2}\ .
\ee
Here $g$ is a universal coupling constant, which
is considered in the K-matrix fit as an unknown parameter, and $\Theta$
is the mixing angle of the
 $n\bar n=(u\bar
u+d\bar d)/\sqrt{2}$ and $s\bar s$ components in $\eta$ and $\eta'$:
$\eta=n\bar n \cos\Theta -s\bar s\sin \Theta $ and $\eta'=n\bar n
\sin\Theta +s\bar s\cos \Theta $, the data give us $\Theta\simeq
38^\circ$ \cite{PDG,mixangle}.

For the decay of scalar $K_0$, the couplings are as  follows:
\begin{eqnarray}
\label{K0decay}
K_0^0\to K^+\pi^-\, &:&\; \frac{g}{2}\ , \quad
K_0^0\to K^0\pi^0 \,:\; -\frac{g}{2\sqrt 2}\ , \nonumber \\
K_0^0\to K^0\eta\,&:&\; \frac{g}{2}(\frac{1}{\sqrt 2}\cos \Theta
-\sqrt\lambda \sin \Theta) \ ,
\nonumber \\
K_0^0\to K^0\eta'\,&:&\; \frac{g}{2}(\frac{1}{\sqrt 2}\sin \Theta
+\sqrt\lambda \cos \Theta)\ .
\end{eqnarray}
The scalar--isoscalar $f_0$-states can be a mixture of
the quark--antiquark and gluonium components,
$ q\bar q\cos\alpha+ gg \sin\alpha \ $,
where the $q\bar q$-state is the mixture of
nonstrange and strange quarks,
 $n\bar n=(u\bar
u+d\bar d)/\sqrt{2}$ and $s\bar s$:
\be
 q\bar q =
n\bar n \cos\varphi+s\bar s \sin\varphi    \; .
\label{3.2}
\ee
For the pure  $q\bar q$ state, the quark combinatorics gives us the
following couplings in the leading terms of the $1/N $-expansion:
\begin{eqnarray}
\label{f0decay}
 f_0(q\bar q)\to \pi^+\pi^-,\, \pi^0\pi^0\,&:&\;
\frac{ g}{\sqrt  2}\cos\varphi \ , \\
 f_0(q\bar q)\to K^+K^-, \, K^0\bar K^0&:&\;
\frac{g}{2} (\sin\varphi+\sqrt{\frac{ \lambda}{2}}\cos\varphi)
\ , \nonumber
\\
f_0(q\bar q)\to \eta\eta &:&\qquad
g\; (\frac{\cos^2\Theta}{\sqrt 2}\cos\varphi
+\sqrt{\lambda}\ \sin\varphi\ \sin^2\Theta)
\ , \nonumber
\\
f_0(q\bar q)\to \eta\eta'&:&\qquad
g\  \sin\Theta\;\cos\Theta\
(\frac{1}{\sqrt 2}\cos\varphi-
\sqrt{\lambda}\;\sin\varphi  )
\ . \nonumber
\end{eqnarray}
The quark combinatorics makes it  possible
to perform the nonet classification of the $q\bar q$
bare states.
For the members of the same $q\bar q$ nonet ($f^{(bare)}_0(1), \
f^{(bare)}_0(2), \ a^{(bare)}_0, \ K^{(bare)}_0$)
this means:
\begin{description}
\item[(1)] The angle difference between
isoscalar nonet partners should be
$90^\circ$:
\be
\varphi [f^{(bare)}_0(1)]-\varphi [f^{(bare)}_0(2)]\simeq
90^\circ \; ,
\label{25}
\ee
that gives us the orthogonality of the flavour wave functions of
$f^{(bare)}_0(1)$ and $f^{(bare)}_0(2)$.
\item[(2)] Coupling constants $g$ presented in Eqs.
(\ref{a0decay}), (\ref{K0decay}) and (\ref{f0decay})
should be roughly
equal to each other for all nonet partners:
\be
g[f^{(bare)}_0(1)]
\simeq g[f^{(bare)}_0(2)] \simeq g[a^{(bare)}_0] \simeq
g[K^{(bare)}_0] \; .
 \label{26}
\ee
\end{description}
The approximate equality in (\ref{25}), (\ref{26}) follows from
the neglect of the next-to-leading terms in (\ref{a0decay}),
(\ref{K0decay}), (\ref{f0decay}).

\subsubsection{Quark combinatoric
relations for the decay couplings of the gluonium state}

For the decay of the pure gluonium state, the planar diagrams
of Fig. 4b type give us:
\begin{eqnarray}
\label{ggdecay}
 f_0(\rm{gluonium})\to \pi^+\pi^-, \
\pi^0\pi^0&:&\qquad \frac{ G}{\sqrt{2+\lambda}}
\ ,
\\
 f_0(\rm{gluonium})\to K^+K^-, \ K^0\bar K^0&:&\qquad
G \sqrt{\frac{\lambda}{2+\lambda}}
\ , \nonumber
\\
f_0(\rm{gluonium})\to \eta\eta &:&\qquad
\frac{G}{ \sqrt{2+\lambda}}\;
 (\cos^2\Theta+ \lambda\sin^2\Theta\ )
\ , \nonumber
\\
f_0(\rm{gluonium})\to \eta\eta'&:&\qquad
G\;
\frac{1-\lambda}{\sqrt{2+\lambda}}
\ . \nonumber
\end{eqnarray}
One may see that the relations between the glueball decay couplings
and those for the $q\bar q$ state with
$\cos\varphi=\sqrt{2/(2+\lambda)}$
coincide. It is due to the fact that the quark--antiquark
component in the glueball, due to (\ref{lambda}), has the
following content \cite{AVA}:
\be   \label{qq-glueball}
(q\bar q)_{glueball}=
n\bar n \sqrt{\frac {2}{2+\lambda}} +s\bar s\sqrt{\frac
{\lambda}{2+\lambda}}\ .
\ee
The appearance of the quark--antiquark component in the glueball
is caused by direct $gg$--$q\bar q$ mixing, which is not suppressed
by the $1/N$-expansion rules  \cite{t'Hooft}, as well as the mixing
related to the decay processes $gg\to mesons \to gg$. So the glueball
is the quarkonium--gluonium composition as follows: \be
gg\cos\gamma+(q\bar q)_{glueball}\sin\gamma\
\ee
with   $\varphi_{glueball}=\tan^{-1}
\sqrt{\lambda/2}\simeq 26^\circ -33^\circ$ for $\lambda\simeq 0.50-0.85$.
Since the ratios of
couplings for the transitions
$gg\to\pi\pi,K\bar K,\eta\eta,\eta\eta'$ are the same for the
quarkonium $(q\bar q)_{glueball}\to\pi\pi,K\bar K,\eta\eta,\eta\eta'$,
we see that the study of
hadronic decays only does not permit to fix the mixing
angle $\gamma$. However, just this equality of ratios allows one to
fix unambiguously the glueball candidates.

\subsubsection{Nonet classification of scalar bare states
in the \boldmath$K$-matrix analysis}

The nonet classification of scalar bare states has been carried out in
\cite{YF,K,AlexSar}, where the mesons were found which are needed to fix
two nonets
 $1^3P_0q\bar q$ and $2^3P_0q\bar q$.
 The decay couplings
to pseudoscalar mesons for nonet states,
\begin{eqnarray}
&& f^{bare}_0(1),\ f^{bare}_0(2)\ \to\ \pi\pi,K\bar K, \eta\eta,\eta\eta'\ ,
 \nonumber\\
&& a^{bare}_0\ \to\ \pi\eta,\ K\bar K\ ,\quad
 K^{bare}_0\ \to\ \pi K,\ \eta K \ ,
\label{baredecay}
\end{eqnarray}
were determined  using for each nonet only
two parameters ($g$ and $\varphi$, see Eqs.
(\ref{a0decay}), (\ref{K0decay}) and (\ref{f0decay})),
while the parameter
$\lambda$ was fixed in the interval $0.50\leq \lambda\leq 0.85 $.

The
constraints (\ref{a0decay}), (\ref{K0decay}) and (\ref{f0decay})
imposed on
the decay couplings (\ref{baredecay}) provided us with the opportunity
to fix unambiguously the states belonging to the basic nonet:
\begin{eqnarray}  \label{firstnonet}
 1^3P_0q\bar q\ : \quad &&
 f^{bare}_0(700\pm100),\ f^{bare}_0(1220\pm40)\ , \nonumber
   \\
 \quad && a^{bare}_0(960\pm30),\
K^{bare}_0 \left(1220^{+50}_{-150}\right)\ .
\end{eqnarray}
The mixing angles for $f^{bare}_0(700)$  and
$f_0^{bare}(1220)$ satisfy the nonet relation (\ref{25}):
\be
\varphi[f_0^{bare}(700)] = -70^\circ \pm 10^\circ \ ,\;\;
\varphi[f_0^{bare}(1220)] = 20^\circ \pm 10^\circ \ .
\label{firstnonetangle}
\ee
Establishing the second nonet, $2^3P_0q\bar q$, is a more complicated
task. The $K$-matrix analysis gives us two scalar--isoscalar bare
states at 1200--1650 MeV, $f^{bare}_0(1230\pm40)$ and
$f_0^{bare}(1580\pm40)$, whose decay couplings (\ref{baredecay}) obey
the relations appropriate to the glueball.

Systematics of bare  $q\bar q$-states on the $(n,M^2)$-plane helps us
to resolve the dilemma about which one of these states is the glueball.
There systematics (discussed in more detail in Section
1.2) definitely tell us that the state $f_0^{bare}(1580\pm40)$,
which is not on the $q\bar q$-trajectory, is an extra one; so
furthermore we accept this state to be the glueball:
\be
0^{++}\ {\rm glueball}: \quad f^{bare}_0(1580\pm40)\ .
\label{glueball}
\ee
Let us stress once again that couplings for the transitions
$ f^{bare}_0(1580\pm 40) \to \pi\pi,K\bar K, \eta\eta,\eta\eta'$
satisfy glueball decay relations (\ref{ggdecay}).

Having  accepted  $f_0^{bare}(1580\pm40)$ to be the non-$q\bar q$-state,
the nonet $2^3P_0q\bar q$ is constructed in a unique way:
\begin{eqnarray}
 2^3P_0q\bar q\ : \quad &&  f^{bare}_0 (1230\pm40),\
f^{bare}_0(1800\pm30)\ , \nonumber
 \label{11'} \\
 \quad && a^{bare}_0(1650\pm50),\ K^{bare}_0\left(1885^{+50}_{-100}
\right)\ ,
\end{eqnarray}
with
\be
\varphi[f_0^{bare}(1230)]= 40^\circ \pm 10^\circ \ ,\quad
\varphi[f_0^{bare}(1800)]=-60^\circ \pm 10^\circ \ .
\label{secondnonetangle}
\ee
After switching on the decay channels, the bare states
turn into real resonances. For
scalar--isoscalar states we have,
after  the onset of decays,  the transformation as follows:
\begin{eqnarray}
&&
f^{bare}_0(700\pm100)\ \longrightarrow\ f_0(980)\ , \nonumber\\
&& f^{bare}_0(1220\pm40)\ \longrightarrow\ f_0(1300)\ , \nonumber\\
&& f^{bare}_0(1230\pm40)\ \longrightarrow\ f_0(1500)\ ,
\nonumber\\
&& f^{bare}_0(1580\pm40)\ \longrightarrow\ f_0(1200-1600)\ ,
\nonumber\\
&& f^{bare}_0(1800\pm40)\ \longrightarrow\ f_0(1750)\ .
\label{12'}
\end{eqnarray}
Just this transformation of the bare scalar--isoscalar states into
 real mesons is shown in Fig. 3.

\begin{figure}[h]
%Fig. 5
\centerline{\epsfig{file=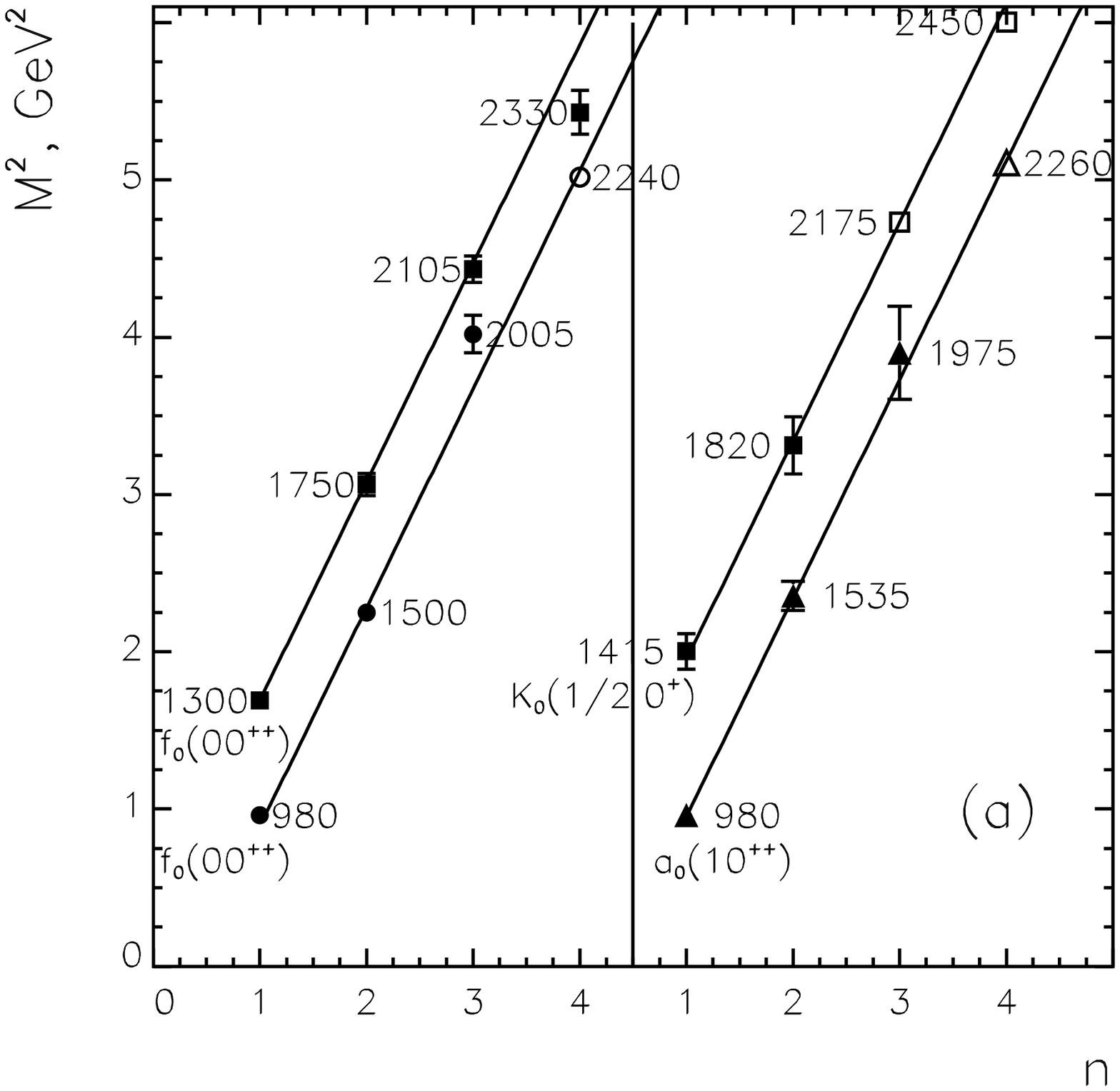,width=6.5cm}
            \epsfig{file=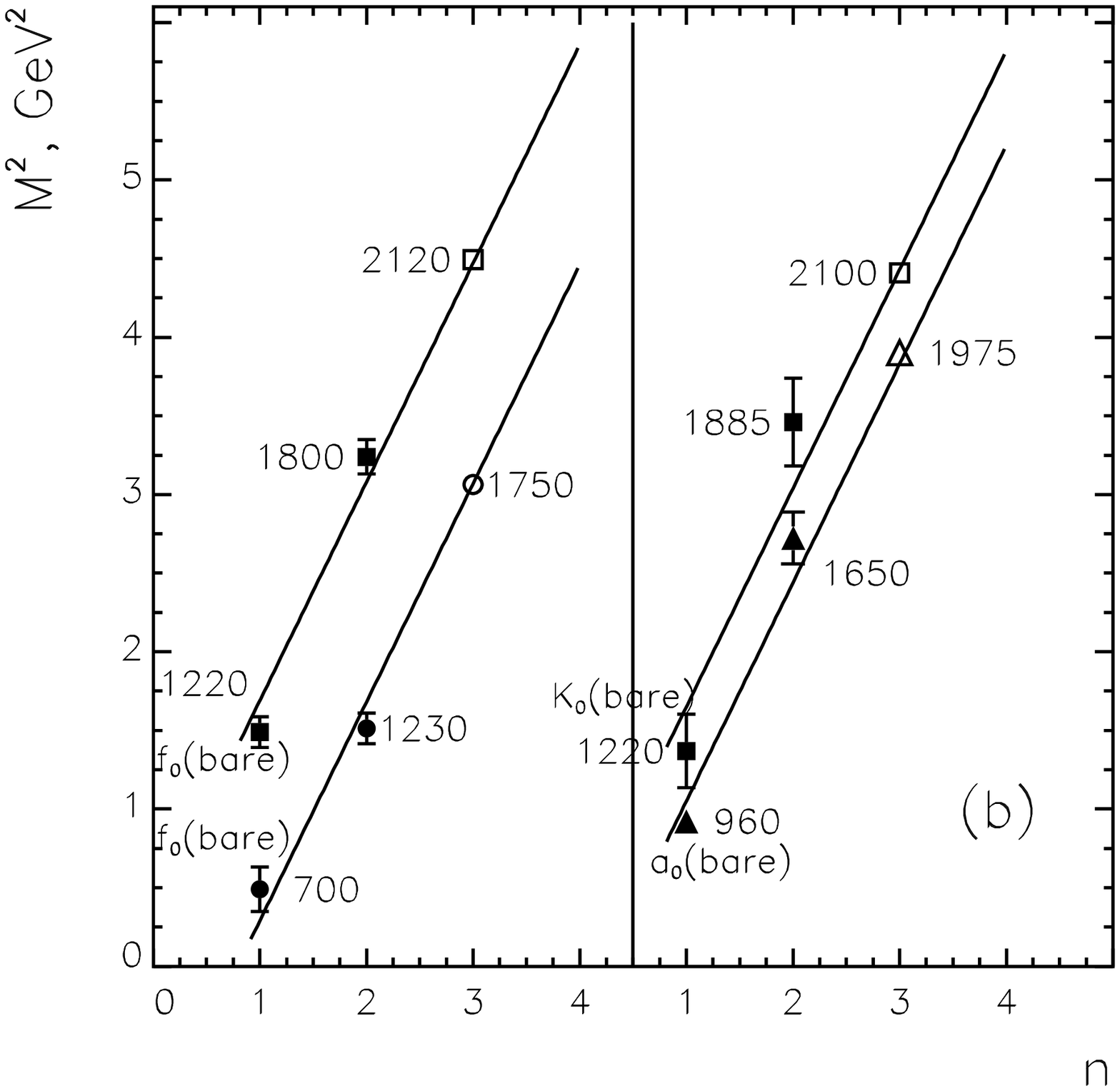,width=6.5cm}}
\caption{ Linear trajectories on the
$(n,M^2)$-plane for scalar resonances (a) and bare scalar states
 (b). Open circles correspond to the predicted states.}
\end{figure}

\subsection{Systematics of scalar states on the
\boldmath$(n,M^2)$-plane}

The systematics of resonances carried out in
\cite{syst} demonstates that all meson resonances can be plotted on
linear trajectories in the $(n,M^2)$-plane,
$M^2=M^2_0+(n-1)\mu^2$,  with a universal slope $\mu^2\simeq 1.3$
GeV$^2$.

This empirical property of $q\bar q$ states may serve us an additional
signature of the $q\bar q$ origin of resonances
$f_0(980)$, $f_0(1300)$, $f_0(1500)$, $f_0(1750)$.
These resonances
fit well to linear
trajectories, with the slope $\mu^2\simeq 1.3$ GeV$^2$. Figure 5a
demonstrates the $(n,M^2)$-trajectories for resonance states with
$00^{++}$, $10^{++}$ and $\frac12 0^{++}$, if the $f_0(1200-1600)$ is
accepted to be the glueball. Note that a doubling of the
$f_0$-trajectories occurs due to the existence of two components,
$n\bar n$ and $s\bar s$.  Similar trajectories for bare states are
shown in Fig. 5b, if the $f_0^{bare}(1580)$ is the gluonium.  The
trajectory slopes for real and bare states almost coincide.

In the variant where $f_0^{bare}(1230)$ is the gluonium, while
the $f_0^{bare}(1580)$ is a $q\bar q$ state,
the linearity of trajectories
is completely broken.
Correct systematics of the lowest scalar $q\bar q$ states is the basis
for a trustworthy search for the glueballs.  Precisely these systematics
allowed us to identify the broad $f_0(1200-1600)$ state as the lowest
scalar glueball. Now we have a strong indication fir the existence of
the tensor glueball \cite{11}, thus opening a new page of physics ---
the glueball physics.

\subsection{The  \boldmath$f_0(980)$ and
$a_0(980)$: are they the quark-antiquark states? }

The nature of mesons $f_0(980)$ and $a_0(980)$ is of principal meaning
for the systematics of scalar states and the search for exotic mesons.
This is precisely why,  till now, theres is a lively discussion  of the
problem of whether the mesons $f_0(980)$ and $a_0(980)$ are the
lightest scalar quark--antiquark particles or whether they are exotics,
like four-quark ($q\bar q q\bar q$) states \cite{Jaffe}, $K\bar K$
molecule \cite{Isgur} or minions \cite{Close}. An opposite opinion
favouring  the $q\bar q $ structure of  $f_0(980)$ and $a_0(980)$ was
expressed in \cite{Narison,Minkowski}.

In Sections 1.1 and 1.2, on the basis of the K-matrix analysis and
systematisation of scalar mesons on the ($n,M^2$)-plane,
 the arguments were discussed
favouring the opinion
that $f_0(980)$ and $a_0(980)$ are dominantly $q\bar q$ states, with a
small ($10-20\%$) admixture of the $K\bar K$ loosely bound component.
There exist other arguments, both qualitative and based on the
calculation of certain reactions, that also support this idea.

First, let
us discuss qualitative arguments.

${\large\bf i)}$ In hadronic reactions, the resonances $f_0(980)$ and
$a_0(980)$ are produced as standard, non-exotic resonances,
with compatible yields and similar distributions. This
phenomenon was observed in
the meson central production at  high energy  hadron--hadron
 collisions (data of GAMS \cite{GAMS-c} and Omega
 \cite{Omega} collaborations) or hadronic decays
of $Z^0$ mesons (OPAL collaboration \cite{OPAL}).

${\large\bf ii)}$ The exotic nature of   $f_0(980)$ and $a_0(980)$
was often argued relying on the surprising proximity of their masses,
while it would be natural to expect the variation of masses in the
nonet to be of the order of 100--200 MeV. Note that the Breit--Wigner
resonance pole, which determines the true mass of the state, is rather
sensitive to a small admixture of hadron components, if the production
threshold for these hadrons lays nearby. As to  $f_0(980)$ and
$a_0(980)$, it is easy to see that a small admixture of the $K\bar K$
component shifts the pole to the $K\bar K$ threshold independently of
whether the pole is above or below threshold. Besides, the peak
observed in the main mode of the
 $f_0(980)$ and $a_0(980)$ decays,  $f_0(980)\to \pi\pi$ and
$a_0(980)\to \eta\pi$, is always slightly  below the $K\bar K$
thresholdl  this mimics a Breit--Wigner resonance with a mass below
1000 MeV ($K\bar K$ threshold). This imitation of a resonance
has created
the legend about a "surprising proximity" of the $f_0(980)$ and
$a_0(980)$ masses.

\begin{figure}[h]
%Fig. 6
\centerline{\epsfig{file=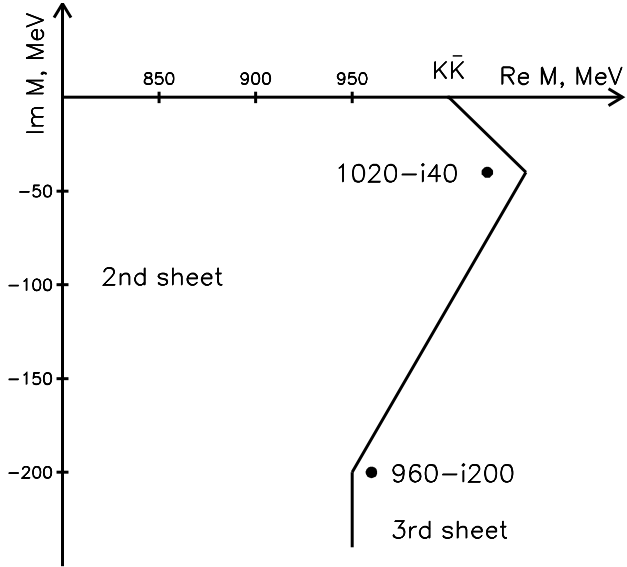,width=9cm}}
\caption{Complex-$M$ plane and location of two poles corresponding to
$f_0(980)$; the cut related to the $K\bar K$ threshold
is shown as a broken line.}
\end{figure}

In fact, the mesons $f_0(980)$ and $a_0(980)$ are characterised not by
one pole, as in the Breit--Wigner case, but two poles (see Fig. 6)
as in the
Flatt\'e formula \cite{Flatte} or $K$-matrix approach; these poles
are rather different for $f_0(980)$ and $a_0(980)$ \cite{YF,K}. Note
that the Flatt\'e formula is unable to give us an adequate description
of spectra near these poles. So we should apply more complicated
representation of the amplitude \cite{content,phidecay} or the
$K$-matrix approach \cite{APS,YF,K},
see also \cite{Au}.

\begin{figure}[h]
%Fig. 7
\centerline{\epsfig{file=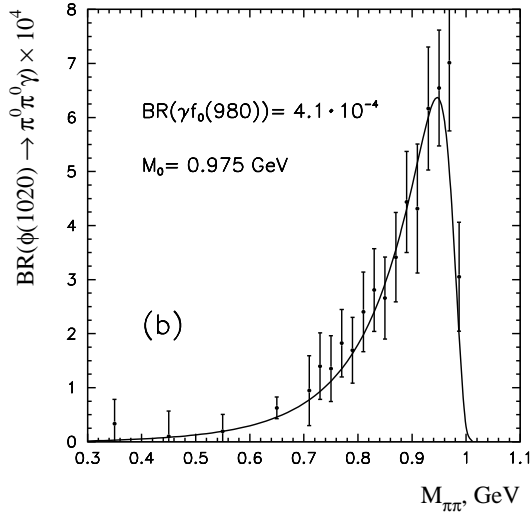,width=0.5\textwidth}}
\vspace{-0.5cm}
\caption{The $\pi\pi$ spectrum from the reaction
$\phi(1020)\to \gamma\pi\pi$ calculated in [35] under the assumption of
the $q\bar q$ nature of $f_0(980)$. Broad tail on the left  (at
$M_{\pi\pi}\sim 700-900$ MeV) is due to the contribution of the second
pole at $M=960-i200$ MeV which is shown in Fig. 6.}
\end{figure}

 In parallel with the above-mentioned qualitative
considerations, there exist convincing arguments which favour the
quark--antiquark nature of $f_0(980)$ and $a_0(980)$:

{\large\bf (I)}
Hadronic decay of the $D^+_s$-meson, $D^+_s\to \pi^+f_0(980)
\to \pi^+\pi^+\pi^-$:
on the quark level, the decay  goes as
$c\bar s \to \pi^+ s\bar s\to  \pi^+f_0(980)$
thus proving the dominance of the $s\bar s$
 component in $f_0(980)$. Our analysis \cite{D+}
showed that
2/3 $s\bar s$ is contained in $f_0(980)$, and this estimate is
supported by the experimental value:
 $BR\left (\pi^+f_0(980)\right)= 57\%\pm 9\%,$
and 1/3 $s\bar s$ is dispersed over the resonances
$f_0(1300)$, $f_0(1500)$, $f_0(1200-1600)$.
    So the reaction $D^+_s \to \pi^+ f_0$ is
a measure of the $1^3P_0s\bar s$ component in the $f_0$ mesons, it
definitely tells us about the dominance of the $s\bar s$ component in
$f_0(980)$, in accordance with results of the $K$-matrix
analysis. The conclusion about dominance of the $s\bar s$ component in
$f_0(980)$  was also made in the analysis
of the decay $D^+_s\to \pi^+\pi^+\pi^-$ in \cite{Gatto,Rupp,Ochs}.

{\large\bf (II)} Radiative decays
  $f_0(980)\to \gamma\gamma$, $a_0(980)\to \gamma\gamma$
 agree well with the
 calculations \cite{9} based on the assumption of the
 quark--antiquark
nature
of these mesons. Let us emphasise again that the calculations favour
the $s\bar s$ dominance in $f_0(980)$.

{\large\bf (III)} Radiative decay $\phi(1020)\to \gamma f_0(980)$
was a subject of lively discussion in the latest years: there existed
an opinion that data on this decay strongly contradict the hypothesis
of  $q\bar q$ nature of $f_0(980)$
\cite{NNA,FEC,DW}.
However, our
calculations \cite{phidecay,9} carried out within both relativistic and
nonrelativistic approaches showed that the $q\bar q$ nature of
$f_0(980)$ agrees well with data, see Fig. 7. In \cite{phidecay}, we
have focused precisely on the applicability of Siegert's
theorem \cite{Sieg} to the decay $\phi(1020)\to \gamma f_0(980)$. We
emphasise that $f_0(980)$ is characterized by two
poles in the complex-mass plane.  Because of that, the notion of mass
difference entering Siegert's theorem is not applicable to
$f_0(700\pm100)$. The use of bare states together with $K$-matrix
approach enables us to describe well experimental data, see Fig. 7.

\begin{figure}[h]
%Fig. 8
\centerline{\epsfig{file=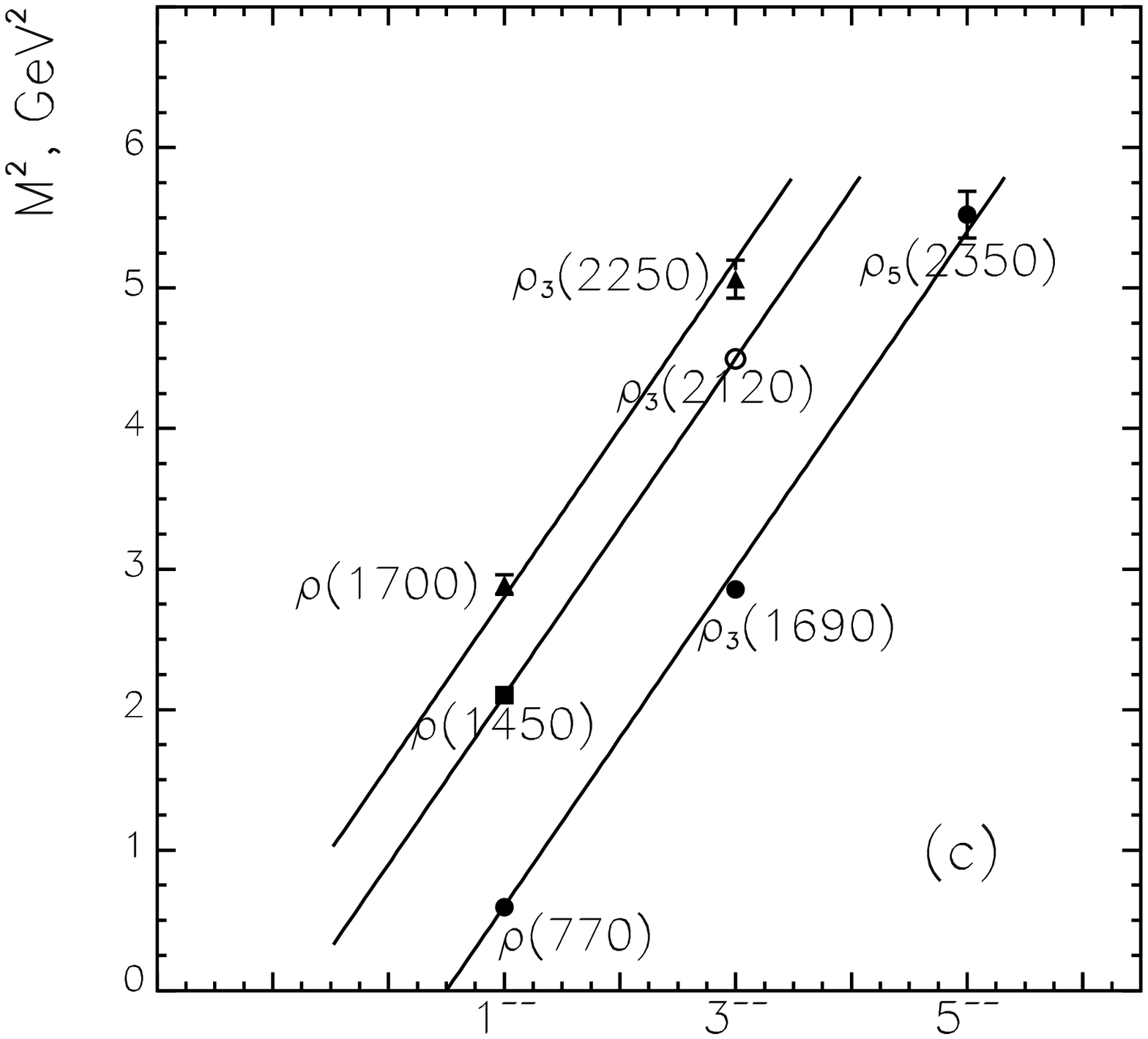,width=5cm}\hspace{-1cm}
            \epsfig{file=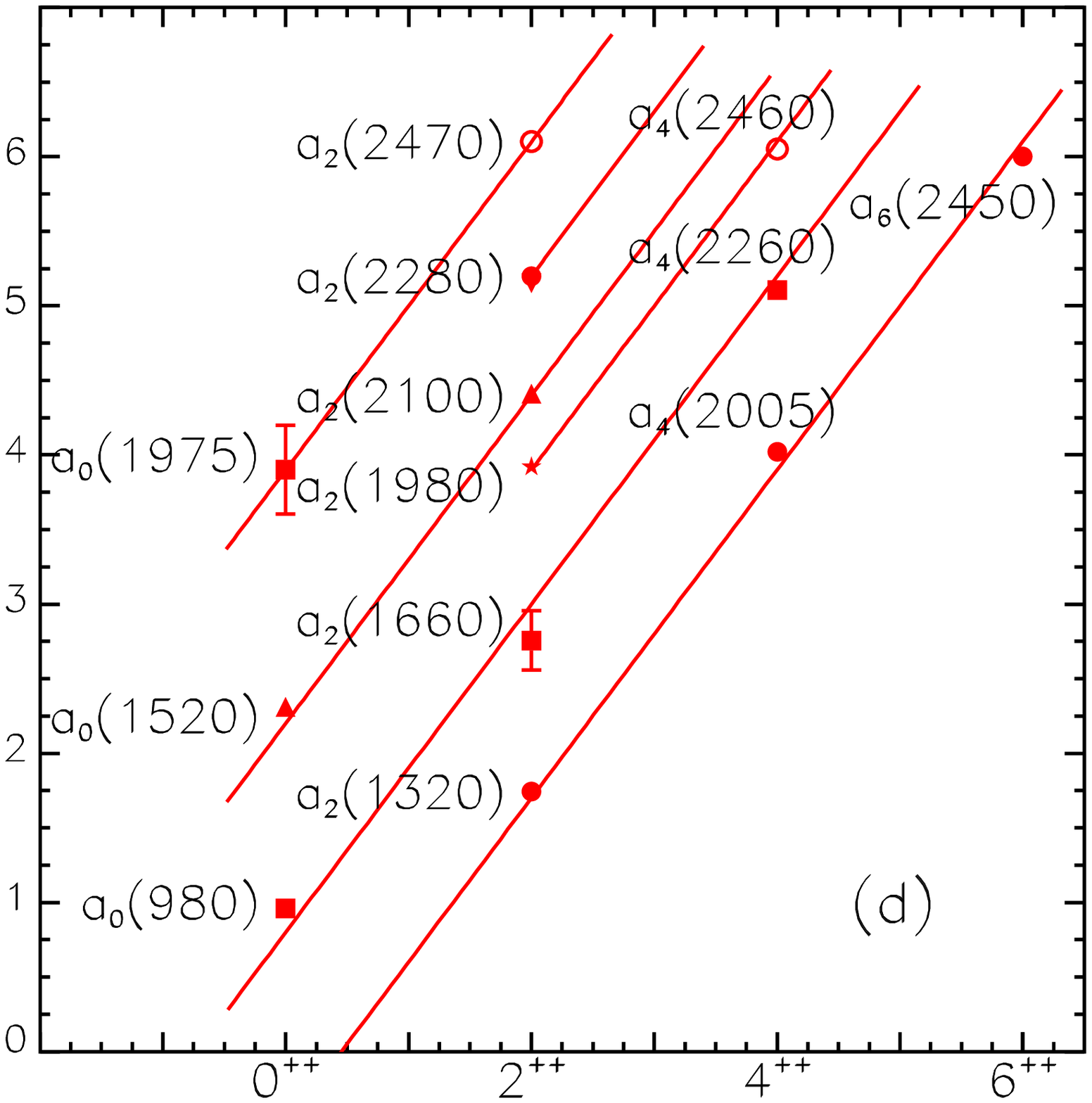,width=5cm}\hspace{-1cm}
            \epsfig{file=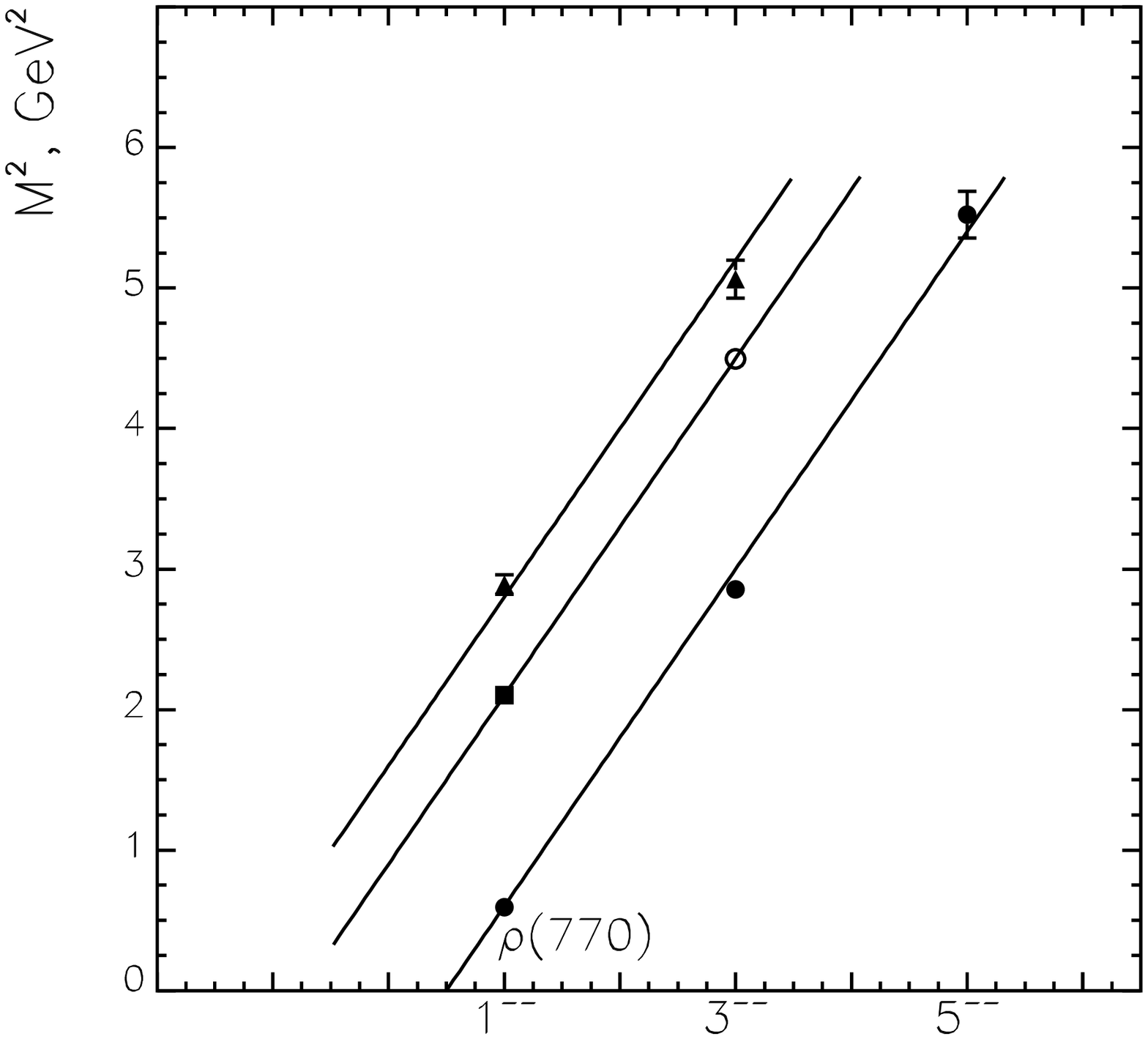,width=5cm}\hspace{-5.15cm}
            \epsfig{file=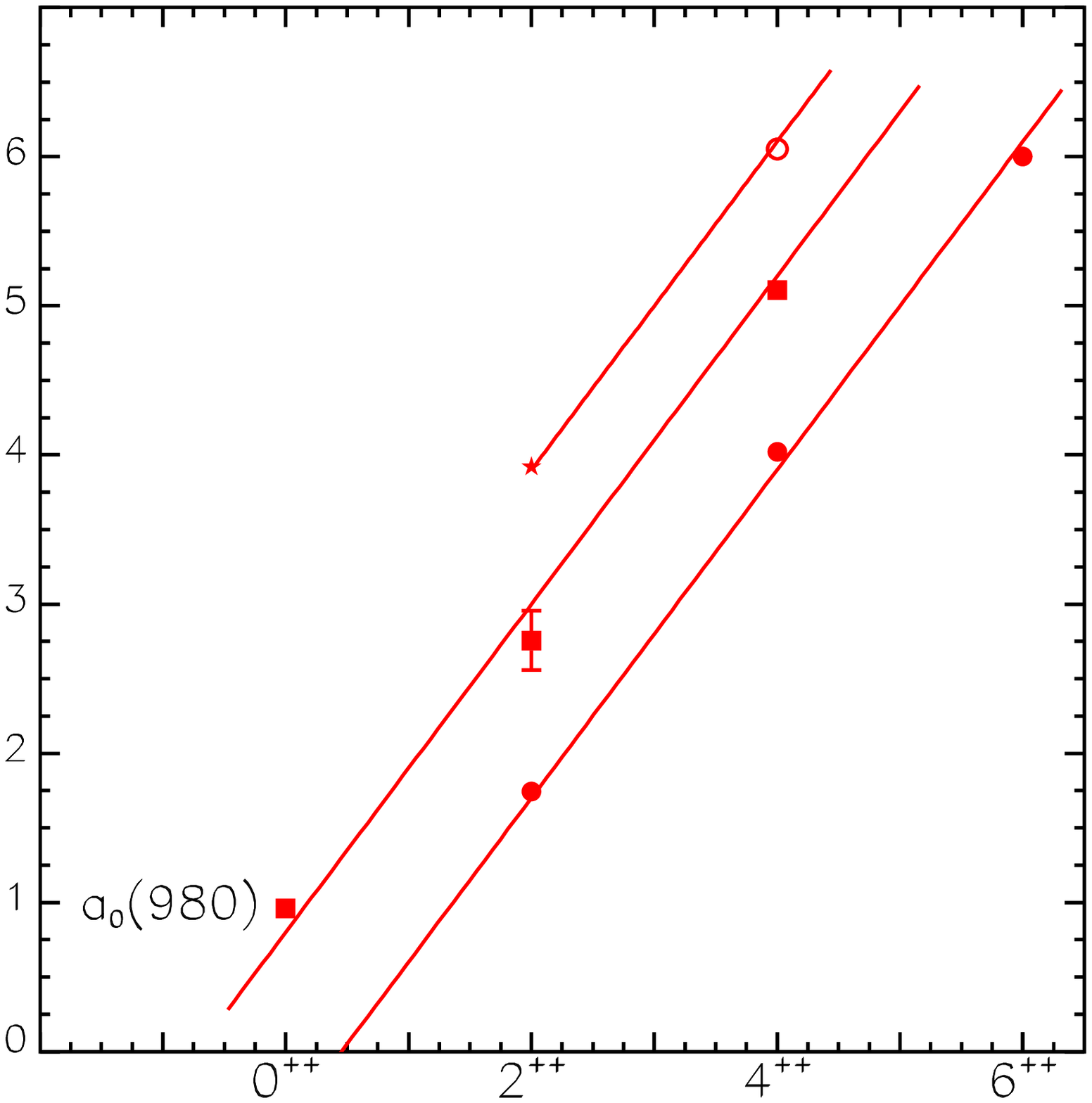,width=5cm}}
\caption{The $\rho_J$ and $a_J$ trajectories on the $(J,M^2)$-plane; the
$a_0(980)$ is on the first daughter trajectory. The right-hand
plot is a combined presentation of $\rho_J$ and $a_J$ trajectories:
if $a_0(980)$ was not a $q\bar
q$ state, there should be another $a_0$ in the mass region
$\sim$1000 MeV.}
\end{figure}

{\large\bf (IV)} Convincing argument in favour of  the $q\bar q$
origin of $a_0(980)$ is given by  considering $(J,M^2)$-planes
for isovector states.  In Fig. 8, the leading and daughter $\rho_J$ and
$a_J$ trajectories are shown. The $a_0(980)$ is located on the
first daughter trajectory. Since the $\rho_J$ and $a_J$ trajectories
are degenerate, the right-hand side $(J,M^2)$-plot demonstrates  the
combined presentation of  low-lying trajectories: one can see that
the $a_0$ state is definitely needed near 1000 MeV.
Had $a_0(980)$ been conisidered as exotics and removed from the
$(J,M^2)$-plane, the $(J,M^2)$-trajectories would definitely demand
another $a_0$ state in this mass region. However, near 1000 MeV we
have only one state, $a_0(980)$.

\subsection{The \boldmath$f_0(1300)$: does it exist?}

In the compilation \cite{PDG}, the resonance $f_0(1300)$ is denoted  as
$f_0(1370)$. It was first observed in the combined analysis of the
reactions $p\bar p(at\ rest,\ liquid\ H_2) \to \pi^0 \pi^0 \pi^0$,
$\eta \pi^0\pi^0$, $\pi^0 \eta \eta$ \cite{f1500},  using also
data on $\pi\pi$-scattering \cite{C-M}. The following
position of the pole was found \cite{f1500}:
\be
         M-i\Gamma/2=(1335 \pm 40)-i(127+30/-20) \; {\rm MeV }\ .
\ee
The K-matrix analysis, which included the  GAMS  \cite{GAMS} and
CERN-M\"unich Collaboration \cite{C-M} data, gave us \cite{YF}:
 \be
         M-i\Gamma/2=(1300 \pm 20)-i(120\pm 20)\; {\rm MeV }\ .
 \ee
The most extensive K-matrix analysis, a combined fit to 16
reactions,  led to the following position of the pole \cite{K}:
 \be
         M-i\Gamma/2=(1315 \pm 20)-i(150\pm 30)\; {\rm MeV }\ .
 \ee
In the reactions $\pi\pi \to \pi\pi$, $\pi\pi \to K\bar K$,
$\pi\pi \to\eta\eta$, the $ f_0(1300)$ is not seen as a peak
because of the strong
interference with the broad state $f_0(1200-1600)$ (or  with a
"background" in the terminology of papers which work with a narrow
mass interval and cannot identify the broad resonance).  The $f_0(1300)$
does not reveal itself as a peak in the reactions
$ p\bar p \to\pi^0 \pi^0 \pi^0$, $\eta \pi^0 \pi^0$, $\pi^0, \eta \eta$
as well, but in these processes it is traced on the Argand-plot:
 with increase of mass, the amplitude creates a circle moving
anticlockwise.

The absence of a visible peak corresponding to $ f_0(1300)$ in the
above-discussed spectra makes some people doubtful about the
existence of this resonance. However, let us emphasise that there exist
reactions where $f_0(1300)$ is precisely seen as a peak, namely,
$\pi^- p \to \pi^+\pi^-n$, $\pi^0\pi^0 n$ at large momentum square
transferred
to the nucleon $|t|\sim 1$ GeV$^2 $ \cite{GAMS,BNL-new}, see Figs. 9
and 10.

\begin{figure}
%Fig. 9
\begin{center}
\epsfig{file=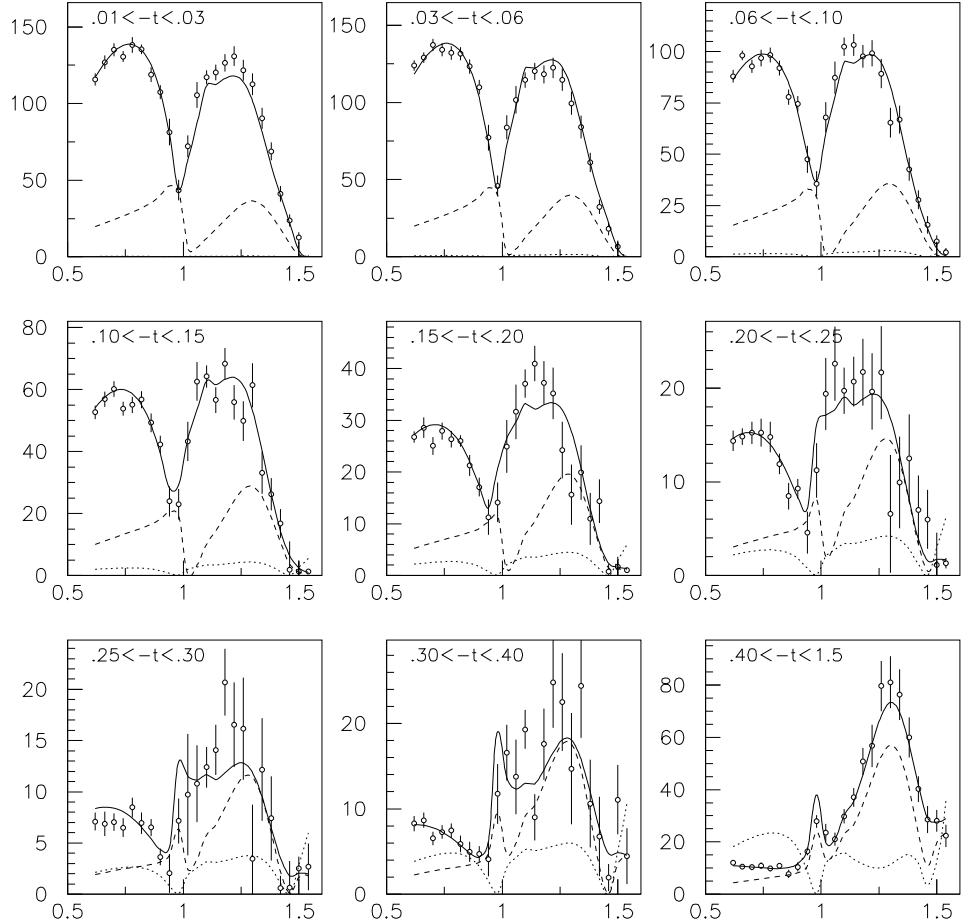,width=14.5cm}
\caption{ Description of the E852 data in
the K-matrix analysis [12](number of events
versus $M_{\pi\pi}$, in GeV)
at different $t$-intervals, in (GeV/c)$^2$.
Dashed curves show contributions from the
$a_{1(leading)}$ reggeon trajectory and dotted
curves from  the $\pi_{(daughter)}$ trajectory;  full
curves correspond to total three-reggeon contributions:
$\pi_{(leading)}+a_{1(leading)}+\pi_{(daughter)}$.}
\end{center}
\end{figure}

\begin{figure}
%Fig. 10
\begin{center}
\epsfig{file=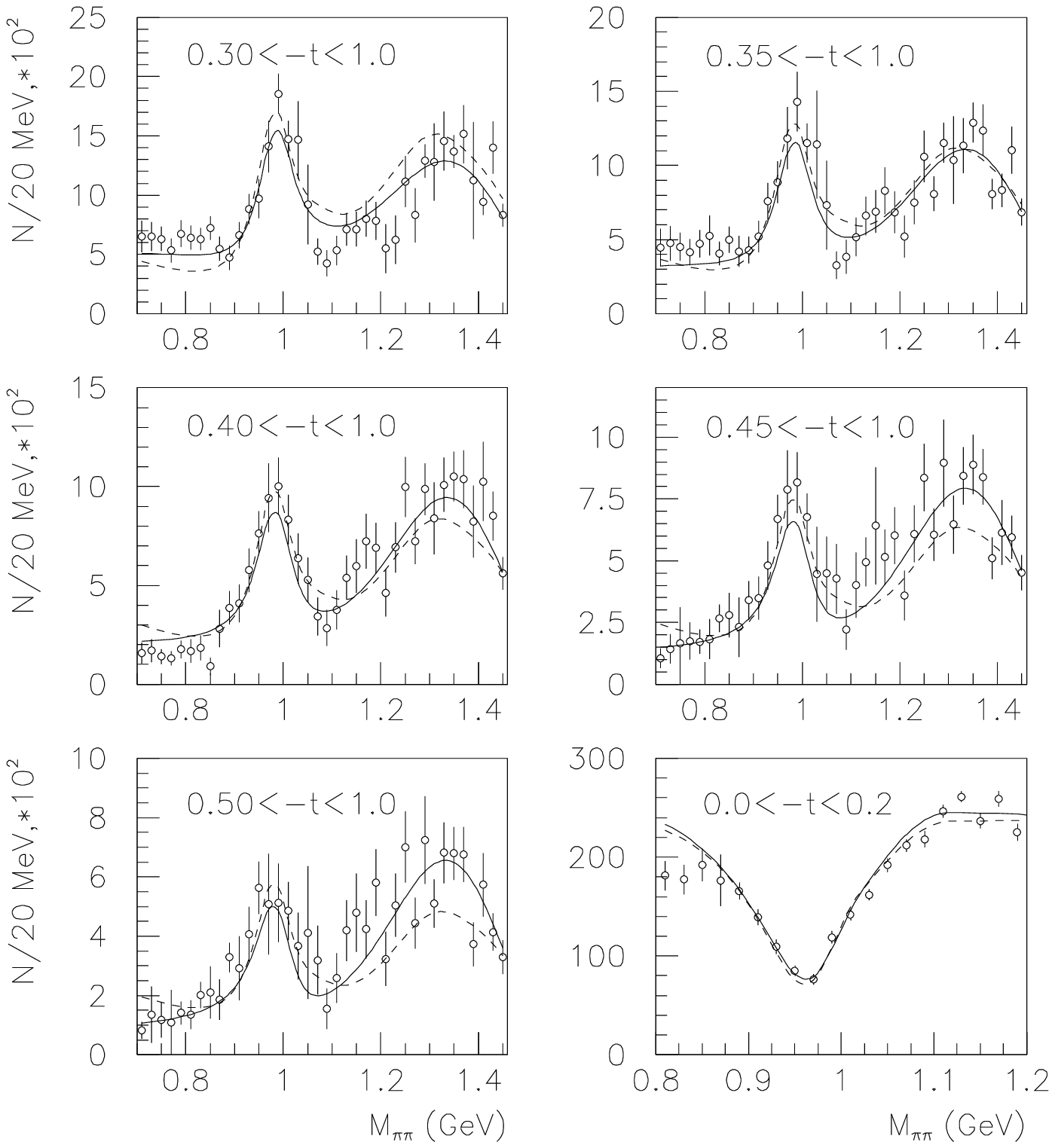,width=14.5cm}
\caption{Results of the
K-matrix analysis [12]: description of the GAMS data at different
$t$-intervals by taking account of the $t$-channel reggeon exchanges
$\pi_{(leading)}+a_{1(leading)}+\pi_{(daughter)}$ -- full curves.
Dashed curves refer to the solution found in previous analysis [11]. }
\end{center}
\end{figure}

At $|t|\sim 1$ GeV$^2 $, the broad state died, while
$f_0(980)$ and $f_0(1300)$ reveal themselves in the $\pi\pi$ spectrum
as clear peaks; Figs. 9, 10 show
data collected in the inerval
$0.4\le |t|\le 1.5$ GeV$^2$
(the data \cite{BNL-new} cover a wider mass
range --- here, on the edge of the spectrum, a small peak is seen at
$M=1500$ MeV; it corresponds to $f_0(1500)$).
Therefore, the experiments \cite{GAMS,BNL-new}  answer the
question of whether $f_0(1300)$ is directly seen in the spectra  as a
peak -- yes, it is.

Now let us survey the situation with the
determination of parameters of $f_0(1300)$ (or $f_0(1370)$ in
the terminology of the PDG \cite{PDG}).

Determination of resonance parameters with the width about 300
MeV can be done in two ways:\\
{\large\bf (i)} In the reaction, if any, with a small background.\\
{\large\bf (ii)}
But if we deal with a resonance
produced in reactions with a considerable background, it is
necessary
to use as many reactions as possible, so as to determine
resonance parameters using the resonance circles on the Argand-plot.
The point is that the position of any resonance in the complex M-plane
is a universal characteristic, while backgrounds may be different
in different processes.  So, precisely the use of a large number of
channels gives us the way to a reliable determination of universal
characteristics.

In \cite{K} the parameters of the $f_0(1300)$ were fixed  in both ways:
in a background-free production reaction  and in reactions with a
considerable background --- the reaction characteristics emerged as
reasonably compatible.

Figures 9 and 10 demonstrate that  a resonance, appearing in
a reaction with large background, can be seen in
the spectrum either as a peak or as a dip, or even as a shoulder, i.e.
it can be simply "invisible".  In the case, when   one--two
reactions  are studied only, one inevitably gets an error in
the mass definition, of the order of $ \Gamma/2$. This is a systematic
error; it depends on the method of fitting and on the background
parameterisation. Concerning the PDG-compilation, we may conclude that
precisely such definitions of mass and width of $f_0(1370)$ have been
collected there. Of course, such a heap of numbers cannot be trusted.
Also it is worth mentioning that averaging over systematic errors is
not a correct procedure.

\section{The \boldmath$\sigma$-meson}

In this Section, a dispersion relation is used to reconstruct
the $\pi\pi$ scattering amplitude in the region of
small $M_{\pi\pi}$; it is then sewn with the $K$-matrix solution
given in \cite{APS,YF,K} for $M_{\pi\pi}\sim 450-1950 $ MeV.
The procedure is as follows.
Data for $\delta^0_0$ are used to construct the N/D amplitude
below 900 MeV; the result is then sewn together with the K-matrix
amplitude, with the purpose of making a continuation to the region
$s=M^2_{\pi\pi}\sim 0$.
In this procedure, we strictly follow the results
obtained for the K-matrix amplitude in the region 450-900 MeV,  where
we may be confident of the results of the K-matrix representation.
Recall that the K-matrix representation allows us to reconstruct
correctly the analytical structure of the amplitude in the region $s >
0$ (threshold and pole singularities) but not the left-hand side
singularities at $s\leq 0$ (singularities related to forces).
Therefore, being cautious, we cannot be quite sure about the validity
of K-matrix results below the $\pi\pi$ threshold.

The
dispersion relation amplitude is found here using the method
suggested in \cite{AKMS,AN}. The
constructed $N/D$-amplitude provides a good description of
$\delta^0_0$ from threshold to 900 MeV, thus including the region
$\delta^0_0 \sim 90^\circ$. This amplitude does not
have a pole at 500--900 MeV; instead, the pole is located
near the $\pi\pi$ threshold.

I discuss the hypothesis that the low-mass pole in the
scalar--isoscalar wave is related to a fundamental phenomenon at large
distances (in hadronic scale).
I argue that the low-mass
pole  corresponds to a white composite particle which is inherent in
subprocesses responsible for  colour confinement forces.

\subsection{Dispersion relation solution for the
\boldmath$\pi\pi$-scattering amplitude below 900 MeV}

The pion--pion  partial scattering amplitude is
 a function of the invariant energy squared,
$s=M_{\pi\pi}^2$. It can be
represented as a ratio $N(s)/D(s)$, where $N(s)$ has a left-hand cut,
due to "forces" (interactions owing to $t$- and
$u$-channel exchanges); the function $D(s)$ is determined by the
rescattering in the $s$-channel. $D(s)$ is given by the
dispersion integral along the right-hand cut in the complex-$s$ plane:
\be
\label{s33}
A(s)=\frac{N(s)}{D(s)}\; , \;\;\;D(s)=1-\int
\limits_{4\mu^2_\pi}^\infty \frac
{ds'}{\pi} \frac{\rho(s')N(s')}{s'-s+i0}\; .
\ee
Here $\rho(s)$ is the invariant $\pi\pi$ phase space, determined as
$\rho(s)=(16\pi)^{-1} \sqrt{1-4\mu^2_{\pi}/s}$.
It was supposed in (\ref{s33}) that $D(s) \to 1$
as $s\to \infty$ and CDD-poles are absent; (a detailed description of
the $N/D$-method may be found in \cite{Chew}).

The $N$-function can be written as an integral along the left-hand
cut as follows:
\be \label{s34}
N(s)=\int
\limits_{-\infty}^{s_L}  \frac{ds'}{\pi}\frac{L(s')}{s'-s}\; ,
\ee
where the value $s_L$ marks the beginning of the
left-hand cut. For example,
for the one-meson exchange diagram $g^2/(m^2 -t)$, the
left-hand cut
starts at $s_L=4\mu_\pi^2-m^2$, and the
$N$-function at this point has a logarithmic singularity; for
the two-pion exchange, $s_L=0$.

Below we work with the amplitude $a(s)$, which is defined as follows:
\be \label{s35}
a(s)= \frac {N(s)}{8\pi \sqrt{s}\left (1-P\int
\limits_{4\mu_\pi^2}^\infty
\frac{ds'}{\pi}\frac{\rho(s')N(s')}{s'-s}\right ) }\; .
\ee

The amplitude $a(s)$  is related to the scattering phase shift:
$a(s)\sqrt{s/4-\mu_\pi^2} = \tan \delta^0_0$.
 In (\ref{s35}), the threshold singularity is explicitly singled out,
 so the function $a(s)$  contains the obly the left-hand cut
plus poles corresponding to zeros
of the denominator of the
right-hand side (\ref{s35}): $1=P\int\limits _{4\mu_\pi^2}^\infty
(ds'/\pi)\cdot \rho(s')N(s')/(s'-s) $. The pole of $a(s)$
at $s>4\mu_\pi^2$ corresponds
to the phase shift
value $\delta^0_0 = 90^\circ$. The phase of the $\pi\pi$
scattering  reaches the value $\delta^0_0 = 90^\circ$ at $\sqrt{s}=
M_{90}\simeq
850$ MeV. Because of that, the amplitude $a(s)$ may be represented in
the form:
\be \label{s36}
a(s)=\int\limits_{-\infty}^{s_L}  \frac{ds'}{\pi}\frac{\alpha(s')}{s'-s}+
\frac{C}{s-M^2_{90}}+D.
\ee
For the reconstruction of the low-mass amplitude, the parameters
$D,C,M_{90}$ and $\alpha(s)$ have been determined by fitting to
experimental data.  In the fit we have used a method which has been
verified in the analysis of low-energy nucleon-nucleon amplitudes
\cite{AKMS}.  Namely, the integral on the right-hand side of
(\ref{s36}) has been substituted by the sum
\be  \label{s37}
\int\limits_{-\infty}^{s_L}
\frac{ds'}{\pi}\frac{\alpha(s')}{s'-s} \to \sum_{n} \frac{\alpha_n}{s_n
-s}\ ,
\ee
with $ -\infty < s_n \leq s_L$.

The description of data using the $N/D$-solution
with six terms in the sum (\ref{s37}) is demonstrated in Fig.
11a. The parameters entering
 the amplitude $a(s)$ are given below:
\be  \nonumber
\begin{tabular}{|c|c|c|c|c|c|c|}
\hline
$s_n\;\mu^{-2}_\pi$ & -9.56 & -10.16 & -10.76 & -32 & -36 & -40 \\
\hline $\alpha_n\;\mu^{-1}_\pi$ & 2.21 & 2.21 & 2.21 & 0.246 & 0.246 &
0.246 \\ \hline \multicolumn{7}{|c|}{$M_{90}=6.228\;\mu_\pi, \qquad
C=-13.64\;\mu_\pi,\qquad D=0.316\;\mu^{-1}_\pi$} \\
\hline
\end{tabular}
\nonumber
\ee
The scattering
length and Adler zero in this solution are equal to
\be  \label{s38}
a^0_0\simeq 0.22\mu_\pi^{-1}\qquad
 s=0.12 \mu_\pi^2 \ .
\ee
 The $N/D$-amplitude is sewn
with the $K$-matrix amplitude of \cite{YF,K}, and Fig. 11b
demonstrates the level of the coincidence of the amplitudes  $a(s)$ for
both solutions; (the values of $a(s)$ which correspond to the $K$-matrix
amplitude are shown with error bars determined in \cite{YF,K}).

\begin{figure}
% Fig. 11
\centerline{\epsfig{file=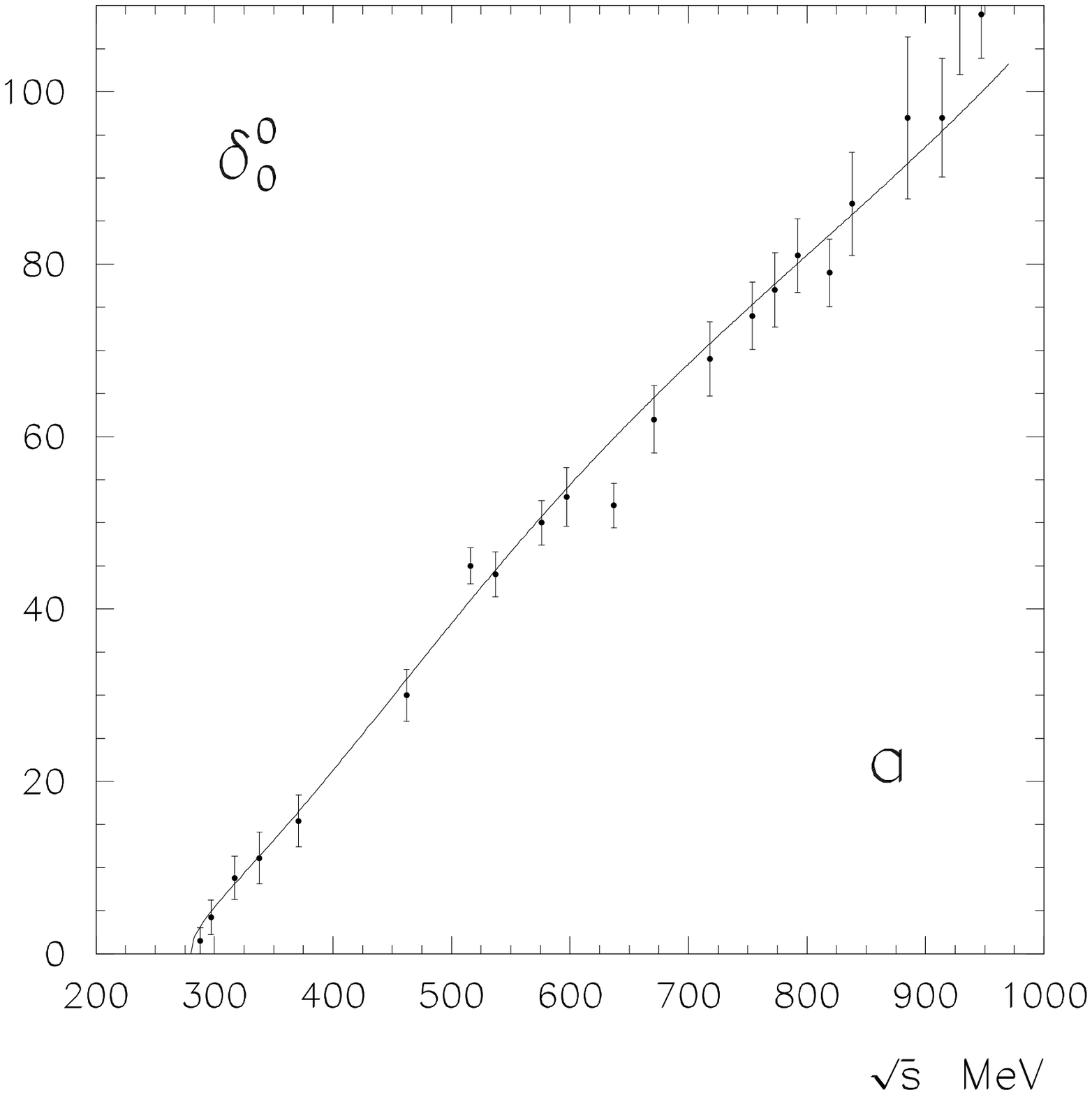,width=7.5cm}\hspace{1cm}
            \epsfig{file=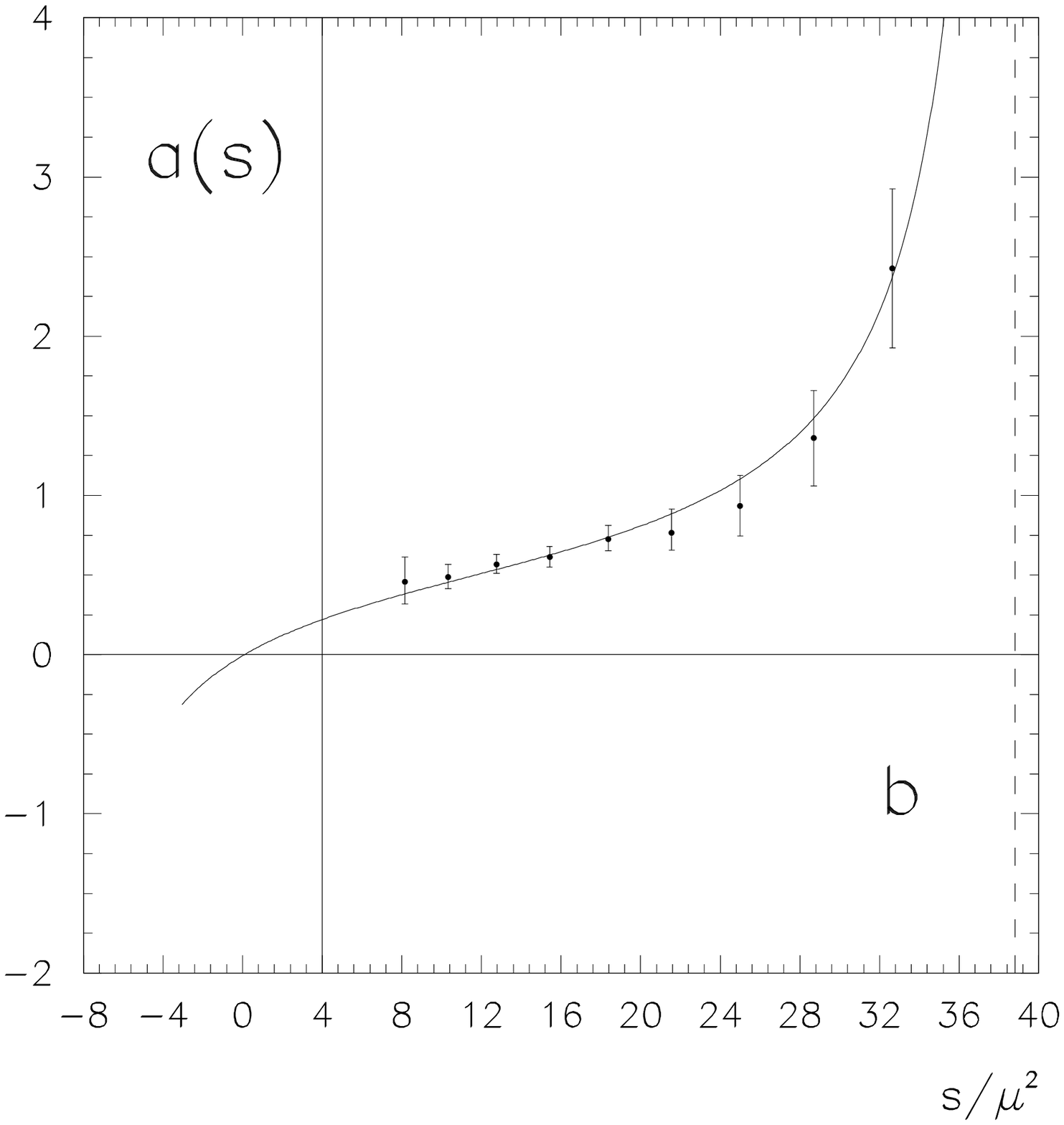,width=7.5cm}}
\caption{a)  Fitting to data on $\delta^0_0$ using the
$N/D$-amplitude.  b) Amplitude $a(s)$ in the $N/D$-solution
(solid curve) and the $K$-matrix approach [4,6] (points with
error bars). }
\end{figure}

\begin{figure}
% Fig. 12
\vspace{3cm}
\centerline{\epsfig{file=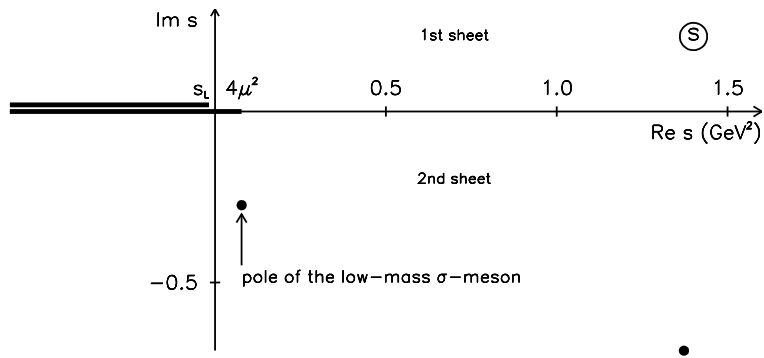,width=10cm}}
\caption{Complex-s plane and singularities of the $N/D$-amplitude}
\end{figure}

The dispersion relation solution has a correct analytical structure at
$|s|<1$ GeV$^2$.  The amplitude has no
poles on the first sheet of the complex-$s$ plane;
the left-hand cut of the $N$-function after
the substitution (\ref{s37})  transforms into a set of
poles on the  negative piece of the real $s$-axis: six poles of the
amplitude (at $s/\mu_{\pi}^2= -5.2,\; -9.6,\; -10.4,\; -31.6,\;
-36.0,\; -40.0$) represent the left-hand singularity of $N(s)$. On the
second sheet (under the $\pi\pi$-cut) the amplitude has two poles:  at
$s\simeq (4-i14)\mu^2_{\pi}$  and $s\simeq (70-i34)\mu^2_{\pi}$ (see
Fig. 12). The second pole,  $s=(70-i34)\mu^2_{\pi}$, is located
beyond the region under consideration, $|s|<1$ GeV$^2$; (nevertheless,
let us emphasise that the $K$-matrix amplitude \cite{YF,K} has a pole
corresponding to the broad state $f(1200-1600)$
just in the region of the second pole of the $N/D$-amplitude).
The pole near the threshold, at
\be \label{s39}
s\simeq (4-i14)\mu^2_{\pi} \ ,
\ee
is precisely what we discuss.
The $N/D$-amplitude has no poles at $Re \sqrt
s \sim 600-900$ MeV, despite the phase shift
$\delta^0_0$ reaches $90^\circ$ here.

The data for $\delta^0_0$ above the $\pi\pi$ threshold do not fix the
$N/D$-amplitude rigidly. The position of the low-mass pole can be
easily varied in the region $Re\; s \sim (0 - 4)\mu_{\pi}^2$, and there
are simultaneous variations of the scattering length in the interval
$a^0_0 \sim (0.21 - 0.28 ) \mu^{-1}_\mu $ and Adler zero at
$s\sim (0-1)\mu_{\pi}^2$. The discussed solution (37) is chosen
because recent data for the scattering length $a^0_0 \sim (0.228
\pm 0.015 ) \mu^{-1}_\mu $
\cite{pislak} coincide with the value presented
in (\ref{s38}).

Let us note that the way of reconstruction of the dispersion relation
amplitude used here differs from the mainstream attempts to
determine the $N/D$-amplitude. In the classic $N/D$ procedure,
that is the bootstrap one; the pion--pion amplitude is to be determined
by analyticity, unitarity and crossing symmetry, giving a unique
determination of the left-hand cut by crossing channels.
However, the bootstrap procedure was not realised till now; the
problems  faced by the present bootstrap
program are discussed in \cite{VV} and references therein.
Nevertheless, one can try to saturate the left-hand cut by
known resonances in the crossed channels. Usually it was supposed
that  dominant contributions to the left-hand cut come from
the $\rho$-meson, $f_2(1275)$ and
$\sigma$ exchanges. Within this scheme, the low-energy amplitude is
restored, being corrected by the available experimental data.

In the scheme used here, see also \cite{AN}, the amplitude
in the physical region at 450-1950 MeV is supposed to be known (the
result of the K-matrix analysis) -- then a continuation of the amplitude
is carried out from the region 450-900 MeV to smaller
masses; this continuation is corrected by the data.  As a result, we
reconstruct the pole near the threshold (the low-mass $\sigma$-meson)
and the left-hand cut (although with lesser accuracy --- on a
qualitative level).

In approaches which take into account
the left-hand cut as a contribution of known
meson exchanges, the following low-mass pole positions were obtained:\\
(i) dispersion relation approach, $s \simeq (0.2-i22.5)\mu_\pi^2 $
\cite{Basdevant}, \\
(ii) meson exchange models, $s \simeq (3.0-i17.8)\mu_\pi^2 $
\cite{Zinn}, $s \simeq (0.5-i13.2)\mu_\pi^2 $ \cite{Bugg},
$s \simeq  (2.9-i11.8)\mu_\pi^2 $ \cite{Speth}.

\subsection{ Low-mass pole as eyewitness of confinement}

It was suggested in \cite{AN,ufn04} that the existence of the light
$\sigma$-meson may be due to a singular behaviour of
forces between quark and antiquark
at large distances;
(in quark models they are conventionally
called "confinement forces"). The scalar confinement
potential, which is needed for the description of
the spectrum of the $q\bar q$-states in the region
1000--2000 MeV, behaves at large hadronic distances as
$V^{(c)}_{confinement}(r)\sim\alpha r$, where
$\alpha\simeq 0.19\,$GeV$^2$. In the momentum representation,
such a growth
of the potential is associated with  singular
behaviour at small  $q$:
\be \label{A}
V^{(c)}_{confinement}(q)\ \sim\ \frac1{q^4}\ .
\ee
In colour space, the main contribution comes from the
component $c=8$, i.e. the confinement forces should
 be the octet ones.   The question that is  crucial
for the structure of the $\sigma$-meson is as follows:
is there a component with the colour singlet
$V^{(1)}_{confinement}(q)$ in the singular
 potential (\ref{A})?

If the singular component with $c=1$ exists, then it must reveal itself
in hadronic channels as well; that is, in the $\pi\pi$-channel. In
hadronic channels, this singularity should not be exactly the same as in
the colour octet ones, because
the hadronic unitarization of the amplitude
(which is absent in the channel with  $c=8$) should modify somehow
the low-energy amplitude. One may believe that, as a result of the
unitarization in the channel  $c=1$, i.e. due to the account of
hadronic rescattering, the singularity of $V^{(1)}_{confinement}(q)$
may appear in the $\pi\pi$-amplitude
 on the second sheet,
being split into several poles. The modelling of the scalar
confinement potential, with the account for the decay of unstable
levels \cite{bs}, confirms the pole splitting. One may believe that this
singularity is  what we call "the light $\sigma$-meson".

Therefore, the main question consists in the following:
does the $V^{(1)}_{confinement}(q^2)$ have the same singular
behaviour as $V^{(8)}_{confinement}(q^2)$?
The observed linearity of the $(n,M^2)$-trajectories, up to the
large-mass region, $M\sim2000-2500$ MeV \cite{syst},  favours
the idea of the universality in the behaviour of potentials
$V^{(1)}_{confinement}$ and $V^{(8)}_{confinement}$
at large $r$, or small $q$. To see
that (for example, in the process
$\gamma^*\to q\bar q$, Fig. 13a)  let us
discuss the colour neutralisation mechanism of outgoing quarks as
a breaking of the gluonic string by newly born $q\bar q$-pairs. At
 large distances, which correspond to
the formation of states with large masses, several new
$q\bar q$-pairs should be formed.
It is natural to suggest that a convolution of the quark--gluon combs
governs the interaction forces of quarks at large distances,
 see Fig. 13b.
The mechanism of
the formation of new $q\bar q$-pairs to neutralise
colour charges does not have a
selected  colour component. In this case, all colour  components
$3\otimes\bar3=1+8$ behave similarly, that is, at small $q^2$
the singlet and octet components of the potential
 are uniformly singular, $V^{(1)}_{confinement}(q^2)\sim
V^{(8)}_{confinement}(q^2) \sim1/q^4$.
This is seen in Fig. 13a.
The quark--gluon ladder ensures the
$t$-channel flow of colour charge $c=3$;  so a quark--antiquark
interaction amplitude is the convolution of ladder diagrams
$3\otimes \bar 3=1+8$ and contains
two equivalent singlet and octet  components.
This points to a
similarity of $V^{(1)}_{confinement}$ and
$V^{(8)}_{confinement}$.

\begin{figure}
%Fig. 13
\centerline{\epsfig{file=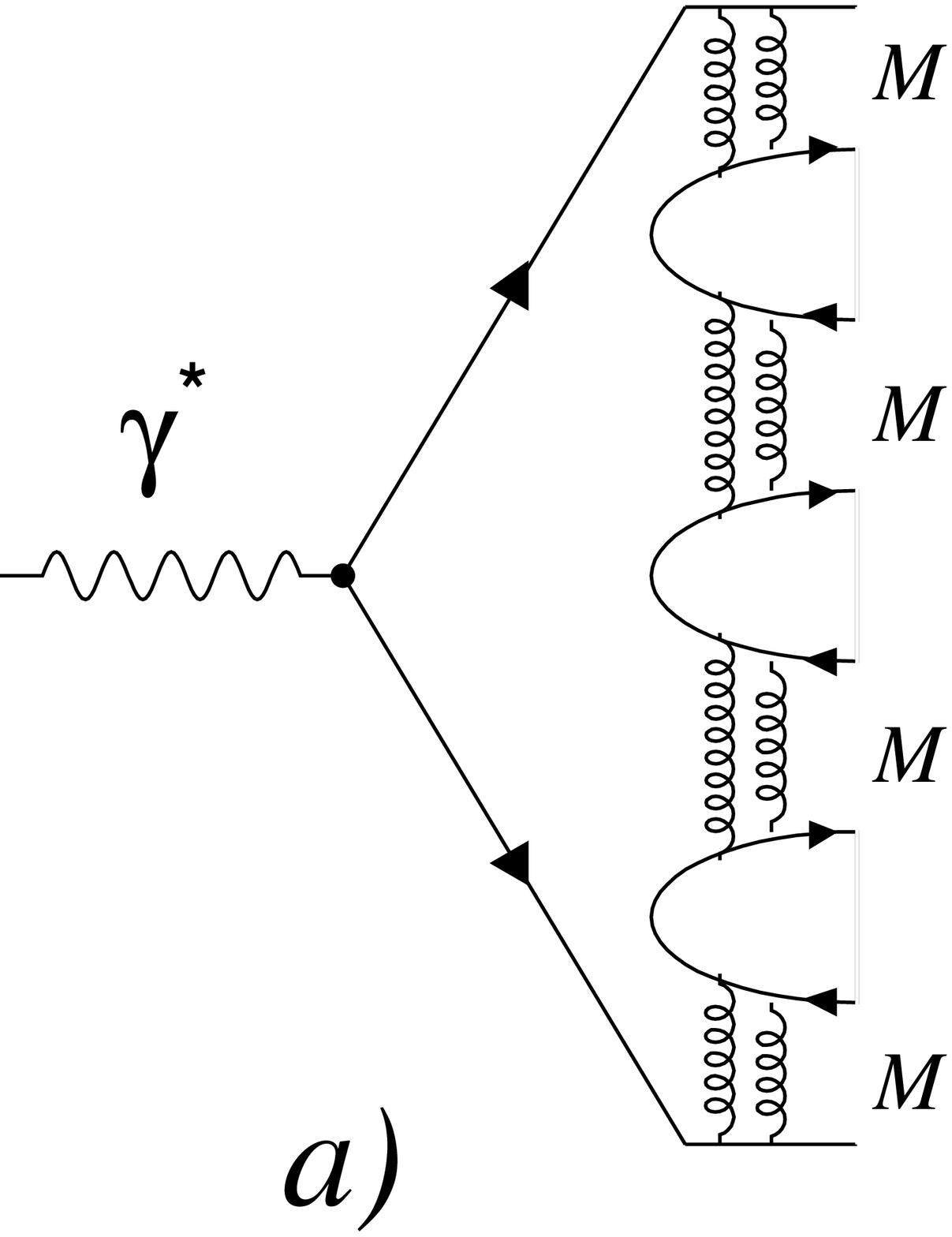,height=4cm}
            \epsfig{file=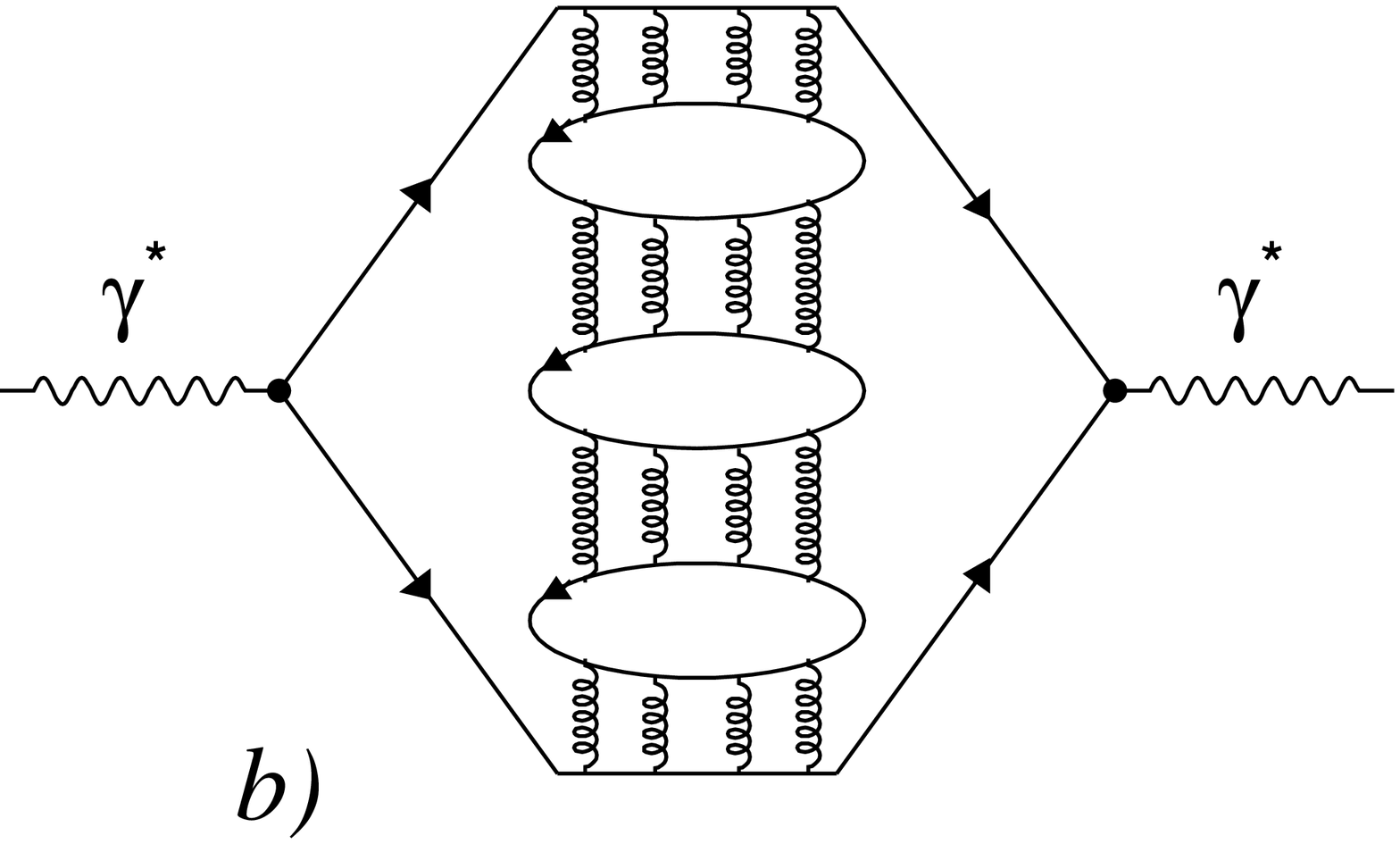,height=4cm}
            \epsfig{file=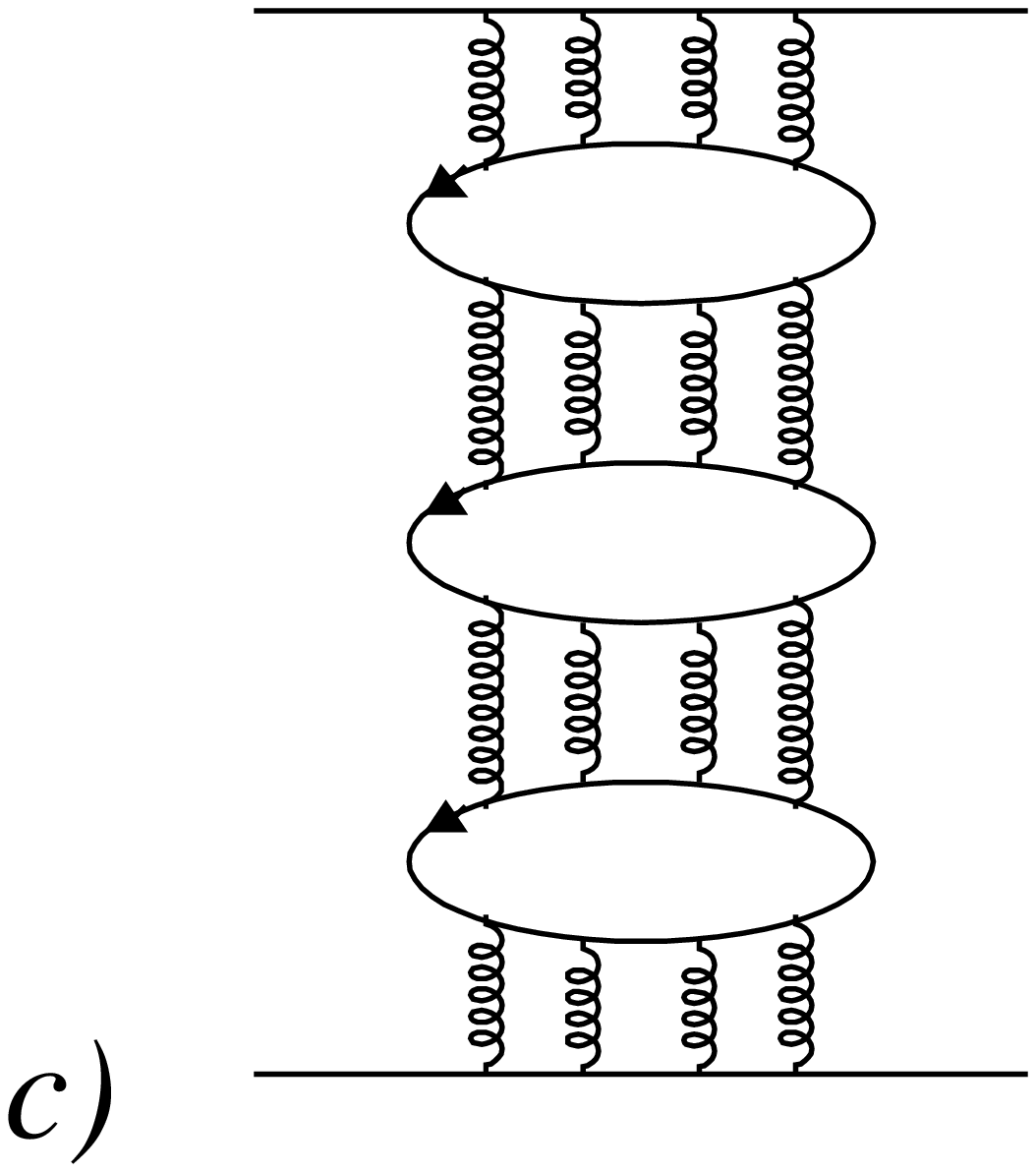,height=4cm}}
\caption{a) Quark--gluonic comb produced by breaking  a string by
quarks flowing out in  the process $e^+e^- \to \gamma^*\to q\bar q\to
mesons$.  b) Convolution of the quark--gluonic combs. c) Example of
diagrams describing interaction forces in the $q\bar q$ systems.}
\end{figure}

\section{Conclusion}

At the present time, one cannot definitely state that lightest $\sigma$
meson exists. The $K$-matrix analysis of meson spectra \cite{K} did
not require the introduction of $\sigma$ meson as an amplitude pole
located near the physical region, though its existence was not denied.
Analytic continuation of the $\pi\pi\to \pi\pi$ amplitude from the
physical region to $s< 4m_\pi^2$ region,  using dispersion
$N/D$ methods, points to the existence of a pole near the $\pi\pi$
threshold. However, such an analytic continuation cannot be a
convincing argument, because of  ambiguities related to the necessity
of accounting for the left-hand cut too. The study within the bootstrap
procedure, with a reconstruction of the left-hand cut contribution at
$0\leq|s|\leq 1$ GeV$^2$, could argue for the existence of the light
$\sigma$ meson; however, one needs to make a sewing with the domain of
the reggeon approach to the amplitude --- such a procedure is still
waiting for its realisation.

In case of the existence of the $\pi\pi$ amplitude singularity near
$s\sim 4m_\pi^2$, one may think that it is due to  "confinement
forces". This singularity can be of the pole type or have a more
complicated structure. It may be of the type of singularity formed by
gluon fields as is shown in Fig. 13b; the inclusion of the quark
loops into gluon planar diagrams is rather natural in the $N/D$
expansion rule --- they do not lead to additional suppression.

The lightest pseudoscalar and scalar $q\bar q$ states are the
 pairs of the $0^{-+}$ and $0^{++}$ mesons as follows:
$\eta(550),\eta'(958)$ and
$f_0^{bare}(700\pm100),f_0^{bare}(1220\pm40)$. In both cases the
lightest particle ($\eta(550)$ for $0^{-+}$ and $f_0^{bare}(700\pm100)$
for $0^{++} $)  is close to a flavour octet; that is their
 common feature. The $f_0^{bare}(700\pm100)$ has undergone a
 noticeable shift of mass in transforming from the bare state to a real
 meson; that is
 due to the decay transitions $f_0\to real\, mesons\to f_0$
 and the proximity of the broad scalar glueball $f_0(1200-1600)$ (or
 maybe the proximity of the $\sigma$ meson). As a result, we observe the
 transformation $f_0^{bare}(700\pm100)\longrightarrow f_0(980)$.

 The role of gluonic states in the formation of lowest $0^{-+}$ mesons
 is not clear. According to \cite{mixangle}, the admixture of gluonium
 component in the $\eta$ meson is small, $\leq 5\%$, while in the
 $\eta'$ meson it is noticeably larger, $\leq 15\%$. There is a variety
 of opinions concerning the mass of the pseudoscalar glueball: it is
 $\sim 2400$ MeV in lattice calculations \cite{lattice}, or
 $\sim  1500$ MeV in string model estimates \cite{Faddeev}, or
$\geq  1700$ MeV, as is obtained from the data on $J/\psi$ radiative
decay, see \cite{BuggR} and references therein. Therefore, it is rather
difficult to say what does it meant by the similarity of flavour
structures in $\eta(550)$ and $f_0^{bare}(700\pm100)$ (both are nearly
flavour octets).

\section*{Acknowledgement}
I thank A.V. Anisovich, Ya.I. Azimov, D.V. Bugg, L.G. Dakhno,
V.A. Nikonov, J. Nyiri, H.R. Petry, A.V. Sarantsev
and V.V. Vereshagin for fruitful discussions.
The article is supported by the RFBR grant N 04--02--17091.

\end{document}